\documentclass[aps,prd,preprint,floatfix,superscriptaddress,nofootinbib,longbibliography,a4paper]{revtex4-1}
\pdfoutput=1
\usepackage{xcolor}
\definecolor{amnd}{RGB}{230,110,0}
\usepackage{graphicx}
\usepackage{textcomp}
\usepackage{orcidlink}
\usepackage{amsmath,amssymb,bm}
\usepackage{booktabs}
\usepackage{placeins}
\usepackage{hyperref}
\usepackage[normalem]{ulem} 
\numberwithin{equation}{section}\renewcommand{\theequation}{\arabic{section}.\arabic{equation}}

\allowdisplaybreaks
\hypersetup{
citecolor=red, colorlinks=true, linkcolor=blue, filecolor=magenta, urlcolor=blue}

\newcommand{\beqa}{\begin{eqnarray}}
\newcommand{\eeqa}{\end{eqnarray}}
\newcommand{\be}{\begin{equation}}
\newcommand{\ee}{\end{equation}}
\newcommand{\ba}{\begin{array}}
\newcommand{\ea}{\end{array}}
\newcommand{\nl}{\nonumber \\}
\newcommand{\eq}{\,&=&\,}
\newcommand{\hc}{\;+\;\mathrm{H.c.}}
\newcommand{\ad}{\,&+&\,}

\newcommand{\GeV}{\mathrm{GeV}}

\newcommand{\chiQCD}{\chi_{\rm QCD}}

\newcommand{\figSolidKey}[1]{\raisebox{0.35ex}{\textcolor[rgb]{#1}{\rule{1.45em}{0.9pt}}}}
\newcommand{\figLongDashKey}[1]{\raisebox{0.35ex}{\makebox[1.65em][l]{\textcolor[rgb]{#1}{\rule{0.65em}{0.9pt}\hspace{0.25em}\rule{0.65em}{0.9pt}}}}}

\newcommand{\figPatchKey}[2]{\raisebox{-0.05ex}{\fcolorbox[rgb]{#1}{#2}{\phantom{\rule{1.15em}{0.55ex}}}}}
\renewcommand{\arraystretch}{1.18}
\begin{document}
\vspace*{0.5cm}
\title{Small-instanton effects in an atlas of Kim-Shifman-Vainshtein-Zakharov axion models}
\bigskip
\author{Ning Chen\orcidlink{0000-0002-0032-9012}}
\email{chenning\_symmetry@nankai.edu.cn}
\affiliation{School of Physics, Nankai University, Tianjin 300071, China}
\author{Saurabh K. Shukla\orcidlink{0000-0001-5344-9889}}
\email{saurabhks@nankai.edu.cn}
\affiliation{School of Physics, Nankai University, Tianjin 300071, China}

\begin{abstract}
\vspace*{1.5cm}
We investigate small-instanton contributions to the axion potential across a range of KSVZ models containing vector-like quarks~(VQs), using naive dimensional analysis. 
We consider scenarios containing a single VQ, multiple identical copies, and sets of distinct VQs, requiring in each case that the gauge couplings remain perturbative up to the Planck scale under the two-loop gauge running. 
The associated fermion zero-mode content varies between these cases, requiring different combinations of mass insertions and scalar--Yukawa contractions for its saturation. 
Increasing the copies of VQs can render the instanton-size integral dominated by instantons of the smallest size, corresponding to the scale near the ultraviolet~(UV) cutoff. 
The resulting contribution then becomes sensitive to the UV completion and the induced potential can compete with, or dominate over, the ordinary QCD contribution. 
Assuming that the QCD and small-instanton potentials are aligned, we determine the resulting axion-mass shift and its consequences for the axion--photon coupling. 
When the small-instanton induced susceptibility becomes comparable to or larger than the QCD susceptibility, the physical axion mass $m_a$ is enhanced at fixed decay constant $f_a$, while the axion--photon coupling $g_{a\gamma\gamma}$ remains controlled by $f_a$ and the anomaly ratio of $E/N$. 
The standard QCD relation among $m_a$, $f_a$ and $g_{a\gamma\gamma}$ is consequently modified, opening new regions of the $(m_a,g_{a\gamma\gamma})$ plane for axion searches.
\end{abstract}

\maketitle
\tableofcontents

\section{Introduction}
\label{sec:introduction}

Small parameters in fundamental physics may reflect dynamical protection by symmetries or environmental selection across a landscape of vacua. 
The cosmological constant provides the canonical example of the latter possibility~\cite{Weinberg:1987dv}, and analogous arguments have been put forward to the electroweak scale through the conditions required for complex nuclei ~\cite{Agrawal:1997gf}. 
The strong-$CP$ angle does not admit an analogous environmental explanation. Nuclear binding and stellar nucleosynthesis remain viable for values many orders of magnitude above the experimental limit~\cite{Ubaldi:2008nf}. 
Known environmental criteria therefore do not account for its observed smallness.

The QCD interaction relevant to the strong-$CP$ problem in the renormalizable Standard Model~(SM) Lagrangian is
\be
\label{eq:theta-term}
{\cal L}\supset
\bar\theta\,\frac{g_s^2}{32\pi^2}
G^a_{\mu\nu}\widetilde G^{ \mu\nu\, a} \,,
\quad
\bar\theta=\theta_{\rm QCD}+\arg\det M_q \,,
\ee
where, $G^a_{\mu\nu}$ is the gluon field-strength tensor, $\widetilde G^{\mu\nu\, a}=\epsilon^{\mu\nu\rho\sigma}G^a_{\rho\sigma}/2$ is its dual, $g_s$ is the QCD gauge coupling, $\theta_{\rm QCD}$ is the QCD vacuum angle, and $M_q$ is the complex quark mass matrix. 
The physical parameter $\bar\theta$ includes the phase transferred between $\theta_{\rm QCD}$ and $M_q$ by anomalous chiral transformations. 
No symmetry within the SM enforces a small value of $\bar\theta$, whereas the neutron electric dipole moment requires $|\bar\theta|\lesssim10^{-10}$~\cite{Abel:2020pzs}. 
This tension defines the strong-$CP$ problem (cf. Refs.~\cite{Kim:1986ax,DiLuzio:2020wdo,Sannino:2026wgx,Williams:2026cec,Williams:2026tuw} for complementary reviews).

Three broad classes of solutions to the strong-$CP$ problem have been put forward in the literature. The Peccei--Quinn~(PQ) mechanism replaces $\bar\theta$ by a dynamical degree of freedom ~\cite{Peccei:1977hh,Peccei:1977ur}, and the associated axion relaxes the effective QCD angle to the $CP$-conserving minimum ~\cite{Weinberg:1977ma,Wilczek:1977pj}. 
Invisible axion realizations introduce either heavy vector-like quarks (VQs), as in Kim-Shifman-Vainshtein-Zakharov~(KSVZ) models ~\cite{Kim:1979if,Shifman:1979if}, or an extended Higgs sector, as in  Dine-Fischler-Srednicki-Zhitnitsky~(DFSZ) models~\cite{Zhitnitsky:1980tq,Dine:1981rt}. 
Further restrictions may arise when the PQ symmetry is tied to flavor or grand unification. 
In flavor models it may also account for the observed fermion hierarchies~\cite{Calibbi:2016hwq}, whereas in unified theories the larger gauge structure can constrain the anomaly coefficients and hence the axion couplings~\cite{Wise:1981ry,Ernst:2018bib,DiLuzio:2020qio,Quevillon:2020aij,Chen:2021haa,Agrawal:2022lsp,Babu:2024udi,Cacciapaglia:2026yvm}. 
A second class imposes $CP$ symmetry on the Lagrangian and breaks it spontaneously, as in Nelson--Barr frameworks, which generate a nonzero CKM phase while maintaining $\bar\theta=0$ at tree level~\cite{Nelson:1983zb,Barr:1984qx,Babu:1989rb,Barr:1991qx,Kuchimanchi:1995rp,Kuchimanchi:2010xs,Kuchimanchi:2023imj,Mohanta:2024wmh} and generate the same at higher order(s). 
The third possibility includes a massless up quark which would instead render $\bar\theta$ unphysical~\cite{Kaplan:1986ru}, but this possibility is excluded by lattice determinations of the light-quark masses ~\cite{FlavourLatticeAveragingGroupFLAG:2021npn}.

 The connection between $\bar\theta$ and the axion potential is non-perturbative. 
 Although $G\widetilde G$ is locally a total derivative term, its spacetime integral need not vanish for gauge configurations with nonzero integer topological charge $Q$. 
 Upon the replacement $\bar\theta\to\bar\theta+a/f_a$, a sector of charge $Q$ enters the path integral with the phase $\exp[iQ(\bar\theta+a/f_a)]$. 
 Summing over topological sectors therefore generates a periodic vacuum energy ~\cite{tHooft:1976snw}. 
 Its curvature at the minimum is the topological susceptibility~$(\chi)$. 
 With the ordinary QCD sector, one has the QCD axion mass~\cite{Weinberg:1977ma,GrillidiCortona:2015jxo,Gorghetto:2018ocs} of 
 \be \label{eq:qcd-axion-mass}
 m_{a,{\rm QCD}}  =\frac{\chiQCD^{1/2} }{ f_a} = \frac{\sqrt{m_u m_d } }{ m_u + m_d } \frac{ m_\pi f_\pi }{f_a} \simeq5.69\,\mu{\rm eV} (10^{12}\,\GeV/f_a) \,.
 \ee 
 The corresponding meV region of this relation is now a common target of axion model building, cosmology, astrophysics and laboratory searches ~\cite{Cicoli:2026fqp}.

The short-distance instantons can contribute to the same periodic axion potential, altering its total curvature, and hence contributing to the standard mass--decay-constant relation in Eq.~\eqref{eq:qcd-axion-mass}. 
Semiclassical estimates of the UV sensitive instanton contribution to the axion potential were computed in Refs.~\cite{Holdom:1982ex,Flynn:1987rs}.
Small-instanton effects have also been examined in factored gauge theories and in models containing additional heavy-fermion zero modes~\cite{Agrawal:2017ksf,Agrawal:2017evu,Cerdeno:2018dqk}.
Partially broken gauge theories were analyzed in Ref.~\cite{Csaki:2019vte}, and the consequences for axion masses and $CP$-violating observables were studied in Refs.~\cite{Bedi:2022qrd,Kivel:2022emq}. 
Our calculation follows the instanton naive dimensional analysis~(NDA) prescription of~\cite{Csaki:2023ziz}, which provides the power-counting rules for fermion zero-mode saturation, the associated loop factors, and the endpoint behavior of the instanton-size integral. 
One-instanton amplitudes have also been studied using functional methods, with applications to string--wall dynamics~\cite{Sesma:2024tcd,Hor:2025gxo}.

A relative phase between the QCD and UV potentials displaces their minima and induces a nonzero effective strong-$CP$ phase. 
UV dynamics that enhances the axion mass can likewise amplify the response to $CP$-violating operators and aggravate the axion-quality problem ~\cite{Bedi:2022qrd,Csaki:2023ziz}. 
Protecting the PQ symmetry from additional UV effects has consequently motivated gauge-protected unified constructions and extra-dimensional realizations of the axion~\cite{DiLuzio:2020qio,Babu:2024udi,Reece:2025thc}.
Throughout this work, the QCD and UV potentials are assumed to have equal periodicity and aligned minima. 
Under this assumption, the contributions due to small-instanton modifies the curvature without shifting the $CP$-conserving vacuum.

In this article, we consider the KSVZ realization of the invisible axion, where the SM is extended by a complex PQ scalar $\Phi$ and some heavy VQs. 
In such minimal KSVZ setup, the SM fermions are PQ neutral and the VQs do not mix with the SM fermions. We use the VQ representations classified in Ref.~\cite{DiLuzio:2016sbl,DiLuzio:2017pfr,Plakkot:2021xyx,Cheek:2023fht,DiLuzio:2024xnt}, and do not include their possible interactions with SM fermions~\cite{Alonso-Alvarez:2023wig,Palavric:2026vej}. 
Few of these VQ representations satisfy the gauge-perturbativity condition adopted in this work. 
We determine the allowed zero-mode closures for all different and the corresponding small-instanton induced susceptibility for one VQ, identical VQ copies, and sets of distinct representations, by using the NDA rules of Ref.~\cite{Csaki:2023ziz}. 
Accordingly, we then study the implications on the axion mass and the axion--photon coupling.

This work is organized as follows. 
Sec.~\ref{sec:instanton-estimates} reviews the semiclassical instanton calculation, fermion zero-mode counting and the NDA estimates. Sec.~\ref{sec:vq-models} constructs an atlas of KSVZ spectra containing one VQ, identical copies or distinct representations, subject to the gauge perturbativity and discusses the UV sensitive instanton contribution to the axion potential. 
Sec.~\ref{sec:axion-pheno} discusses the consequences for the axion mass and axion--photon coupling, and we conclude in Sec.~\ref{sec:summary}. 
The representation-dependent invariants and determinant prefactors entering the instanton estimates are collected in Appendix~\ref{sec:appendix}.

\section{Semiclassical instantons and NDA power counting}
\label{sec:instanton-estimates}

The Euclidean action of an ${\rm SU}(N_c)$ gauge theory coupled to the axion is given by
\be
\label{eq:euclidean-action}
S_E =
\int d^4x\left[
\frac{1}{2}
{\rm Tr}\!\left(G_{\mu\nu}G^{\mu\nu}\right)
-\frac{i g_s^2}{16\pi^2}
\left(\bar{\theta}+\frac{a}{f_a}\right)
{\rm Tr}\!\left(G_{\mu\nu}\widetilde{G}^{\mu\nu} \right)
\right] \,,
\ee
with, $G_{\mu\nu}=G_{\mu\nu}^a T^a$ is the canonically normalized field strength, with $G_{\mu\nu}^a=\partial_\mu A_\nu^a-\partial_\nu A_\mu^a+g_s f^{abc}A_\mu^bA_\nu^c$, and its dual is $\widetilde{G}^{\mu\nu}=\frac{1}{2}\epsilon^{\mu\nu\rho\sigma}G_{\rho\sigma}$. 
Locally, the topological density is a total derivative of
\be
\label{eq:topological-total-derivative}
\frac{g_s^2}{16\pi^2}
{\rm Tr}\!\left(G_{\mu\nu}\widetilde{G}^{\mu\nu}\right)
=\partial_\mu K^\mu \,,
\ee
where $K^\mu= \frac{g_s^2}{16\pi^2} \epsilon^{\mu \nu \rho \sigma} \left( A_\nu^a \partial_\rho A_\sigma^a + \frac{g_s}{3} f^{abc} A_\nu^a A_\rho^b A_\sigma^c \right)$ is the Chern--Simons current. 
The finite-action configurations approach a pure gauge at infinity, and the surface term distinguishes sectors of different winding number. 
The topological charge is defined as
\be
\label{eq:topological-charge}
Q =
\frac{g_s^2}{16\pi^2}
\int d^4x\,
{\rm Tr}\!\left(G_{\mu\nu}\widetilde{G}^{\mu\nu}\right)
\in {\mathbb Z} \,.
\ee
The physical vacuum is consequently a coherent sum of the topological vacua $|n\rangle$
\be
\label{eq:theta-vacuum}
|\theta\rangle =
\sum_{n\in{\mathbb Z}}e^{in\theta}|n\rangle \,.
\ee

The Euclidean equations of motion possess finite-action solutions that interpolate between the neighboring topological sectors. 
By using the positivity of $ {\rm Tr}(G_{\mu\nu}\mp\widetilde{G}_{\mu\nu})^2$, one obtains
\be
\label{eq:instanton-action-bound}
S_E\geq
\frac{8\pi^2}{g_s^2}|Q|
-iQ\left(\bar{\theta}+\frac{a}{f_a}\right)\,.
\ee
For $Q=1$, the bound is saturated by the self-dual Belavin--Polyakov--Schwarz--Tyupkin (BPST) ${\rm SU}(2)$ instanton embedded in ${\rm SU}(N_c)$~\cite{Belavin:1975fg}
\be
\label{eq:self-duality}
G_{\mu\nu}=\widetilde{G}_{\mu\nu}\,,
\quad Q=1\,. 
\ee
The anti-instanton is anti-self-dual and carries $Q=-1$. 
For this configuration, the classical one-instanton action is given by
\be
\label{eq:classical-instanton-action}
S_I=
\frac{8\pi^2}{g_s^2}
-i\left(\bar{\theta}+\frac{a}{f_a}\right)
=
\frac{2\pi}{\alpha_s}
-i\left(\bar{\theta}+\frac{a}{f_a}\right),
\quad
\alpha_s=\frac{g_s^2}{4\pi} \,.
\ee

A unit-charge instanton has bosonic collective coordinates associated with translations, scale transformations and gauge rotations, which are parametrized by its center $x_0$, size $\rho$, and $4N_c-5$ independent gauge orientations. 
An instanton of size $\rho$ probes momenta of order $1/\rho$. 
Quantum fluctuations about the saddle point are controlled when its action is large
\be
\label{eq:semiclassical-domain}
\frac{2\pi}{\alpha_s(1/\rho)}\gg 1 \,,
\ee
while the gauge, Yukawa and scalar couplings entering the zero-mode closures must remain perturbative at the same scale. 
Denoting by $M_{\rm UV}$ the UV cutoff and by $\Lambda$ the matching scale at which the active particle content changes, the instanton-size integral is evaluated over the range of
\be
\label{eq:instanton-size-window}
\frac{1}{M_{\rm UV}}<\rho<\frac{1}{\Lambda} \,. 
\ee
Integrating over the collective coordinates and nonzero quantum fluctuations gives the one-instanton measure of
\be
\label{eq:one-instanton-measure}
d\mu_I =
C_{N_c}\,d^4x\,\frac{d\rho}{\rho^5}
\left(\frac{2\pi}{\alpha_s(\mu)}\right)^{2N_c}
\exp\!\left[-\frac{2\pi}{\alpha_s(\mu)}\right]
(\mu\rho)^{b_3} \,. 
\ee
The factor of $C_{N_c}$~\cite{tHooft:1976snw,Bernard:1979qt} contains the finite part of the nonzero-mode determinants and the gauge-orientation integral. 
The power $2N_c$ arises from the normalization of the bosonic zero modes, while $d\rho/\rho^5$ follows from the scale collective coordinate and dimensional analysis. 
At a fixed instanton size $\rho$, the one-loop QCD $\beta$-function coefficient $b_3$ receives contributions from the fields that are active at the characteristic scale $1/\rho$ as
\be
\label{eq:one-loop-beta-coefficient}
b_3=
\frac{11}{3}N_c
-\frac{2}{3}
\sum_{f}T(R_f)
-\frac{1}{3}
\sum_{s}T(R_s)\,,
\ee
where the sums run over active Weyl fermions and complex scalars~\cite{Sartore:2020gou}.

The integration over the nonzero fluctuations generates the factor $(\mu\rho)^{b_3}$ in the instanton measure. 
At one loop, this factor combines with the classical exponential so that the gauge coupling is effectively evaluated at the characteristic instanton scale $1/\rho$. 
In an interval where the active field content, and hence $b_3$, is constant, this combination can be expressed in terms of the RG-invariant scale $\Lambda_G$ as
\be
\label{eq:running-instanton-action}
\begin{split}
\Lambda_G^{\,b_3}
&\equiv
\mu^{b_3}
\exp\!\left[-\frac{2\pi}{\alpha_s(\mu)}\right] \,,
\\
\exp\!\left[-\frac{2\pi}{\alpha_s(\mu)}\right]
(\mu\rho)^{b_3 }
&=
\exp\!\left[-\frac{2\pi}{\alpha_s(1/\rho)}\right]
=
(\Lambda_G\rho)^{b_3 }\,.
\end{split}
\ee
Across a particle-mass threshold, the active field content changes and hence so does $b_3$. The value of $\Lambda_G$ on one side of the threshold can be determined from its value on the other side by requiring $\alpha_s$ to be continuous at the matching scale by using Eq.~\eqref{eq:running-instanton-action}.

\begin{figure*}[t!]
\centering
\includegraphics[width=0.62\textwidth]{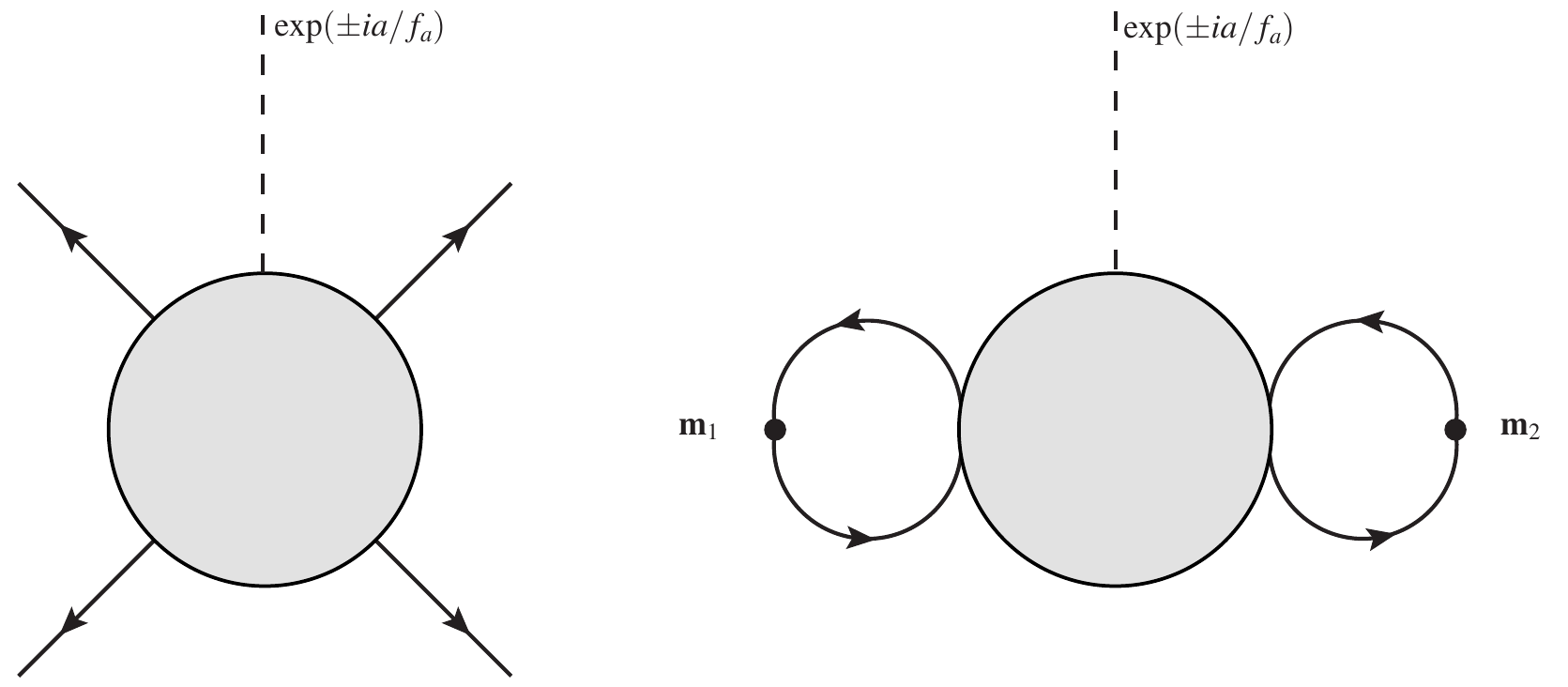}
\caption{'t~Hooft vertex for two vector-like fundamental fermions. 
The diagram in the left panel shows the four fermion zero modes and the dashed axion phase, where the right panel shows their pairwise saturation by the mass insertions $m_1$ and $m_2$, giving $(m_1\rho)(m_2\rho)$. }
\label{fig:thooft-zero-mode-closure}
\end{figure*}

The index theorem~\cite{Vandoren:2008xg} gives the number of zero modes of a left-handed Weyl fermion in a representation $R_f$ as follows~\footnote{Throughout this article, $T(R)$ denotes the Dynkin index of a representation $R$, defined by ${\rm Tr}(T^aT^b)=T(R)\delta^{ab}$, while ${\rm dim}(R)$ denotes its dimension. 
We use the normalization $T({\bf N})=1/2$ for the fundamental representation of ${\rm SU}(N)$.}
\be
\label{eq:fermion-zero-mode-count}
n_f=2\,T(R_f) \,.
\ee
In a unit-charge instanton background, the fermionic functional integral therefore generates the 't~Hooft vertex~\cite{tHooft:1976snw,Vandoren:2008xg} of
\be
\label{eq:t-hooft-operator}
{\cal O}_I\propto
\exp\!\left[
i\left(\bar{\theta}+\frac{a}{f_a}\right)
\right]
\prod_f\left(\psi_f\right)^{2T(R_f)} \,.
\ee
The open 't-Hooft vertices and a simple mass-insertion closure, drawn using JaxoDraw~\cite{Binosi:2003yf}, are shown in Fig.~\ref{fig:thooft-zero-mode-closure}.
A vacuum amplitude is nonzero only after every fermion zero mode has been saturated, and this may be achieved in several ways. 
For example, a mass insertion $m\psi\chi$ absorbs two zero modes~(one-pair) and contributes a factor of 
\be
\label{eq:mass-insertion-factor}
m\rho \,. 
\ee
Four zero modes may instead be saturated by two Yukawa vertices connected by a scalar propagator. 
For an instanton of size $\rho$, the characteristic momentum is $1/\rho$. 
If the scalar mass satisfies $m_\phi\ll 1/\rho$, the scalar field is an active propagating degree of freedom and the closure contributes the NDA factor of
\be
\label{eq:yukawa-loop-factor}
\frac{y_1y_2}{16\pi^2} \,.
\ee
If $m_\phi\gg 1/\rho$, the scalar is not resolved at the instanton scale. 
It is then removed from the effective theory, and its exchange is represented by the corresponding local higher-dimensional operator. 
Additionally, the loop factors entering in the instanton integral can be estimated according to Ref.~\cite{Csaki:2023ziz} as
\be
\label{eq:nda-loop-factor}
(4\pi)^{
-\left[
n_{\rm zero\; modes}
-2n_{\rm vertices}
+2n_{\rm propagators}
\right]} \,,
\ee
where mass insertions are counted as vertices. 
Eqs.~\eqref{eq:mass-insertion-factor}--\eqref{eq:nda-loop-factor} give the factors associated with the available zero-mode closures. 
They are applied after writing the complete 't~Hooft vertex and identifying how every fermion zero mode is saturated. 
Since the calculation relies on semiclassical and perturbative expansions, the gauge, Yukawa and scalar couplings entering a given instanton estimate must remain perturbative over the relevant range of scales.

A closure containing $r$ mass insertions together with $k$ scalar-Yukawa loops. 
Over the interval $1/M_{\rm UV}\leq\rho\leq1/\Lambda$, its contribution to the axion potential is estimated as
\be
\label{eq:general-instanton-potential}
V_I(a)\sim
C_3
\left(\frac{2\pi}{\alpha_s}\right)^6
\exp\!\left[i\left(\bar\theta+\frac{a}{f_a}\right)\right]
\int_{1/M_{\rm UV}}^{1/\Lambda}
\frac{d\rho}{\rho^5}
(\Lambda_G\rho)^{b_3}
\prod_{i=1}^{r}(m_i\rho)
\prod_{j=1}^k
\left(\frac{y_{j_1} y_{j_2} }{16\pi^2}\right)
\cdots \hc\,.
\ee
The Hermitian-conjugate term represents the anti-instanton contribution. 
For a vector-like pair in $R\oplus\overline R$, the number of conjugate zero-mode pairs is $2T(R)$. 
A mass insertion saturates one pair, whereas a scalar-Yukawa loop saturates two. 
Consequently, the complete saturation requires that~\cite{Csaki:2023ziz}
\be
\label{eq:zero-mode-saturation}
2T(R)=2k+r\,.
\ee
Between $\Lambda$ and $M_{\rm UV}$ the active field content, and hence $b_3$, is fixed, and $\Lambda$ is the threshold at which $b_3$ changes. 
The coupling in the collective-coordinate prefactor is evaluated at $M_{\rm UV}$ for an UV-dominated integral and at $\Lambda$ for an IR-dominated integral. 
After extracting the dimensionless Yukawa factors, the dimensionful part reads
\be
\label{eq:instanton-size-power}
\Lambda_G^{\,b_3}
\prod_{i=1}^{r}m_i
\int_{1/M_{\rm UV}}^{1/\Lambda}
d\rho\,\rho^{b_3 +r-5}.
\ee
The exponent of $\rho$ controlling the two endpoints is
\be
\label{eq:instanton-endpoint-exponent}
\Delta=b_3 +r-4\,.
\ee
The resultant integral is UV dominated for $\Delta<0$, IR dominated for $\Delta>0$, and logarithmic for $\Delta=0$. 
Since scalar-Yukawa closures do not supply the positive powers of $\rho$ associated with mass insertions, each such closure lowers $\Delta$ by two and increases the UV sensitivity. 
Its coefficient remains weighted by the corresponding Yukawa couplings and loop factor. For $M_{\rm UV}\gg\Lambda$, Eq.~\eqref{eq:instanton-size-power} gives \( \int_{1/M_{\rm UV}}^{1/\Lambda} d\rho\,\rho^{\Delta-1} = (\Lambda^{-\Delta}-M_{\rm UV}^{-\Delta})/\Delta \). 
By combining the instanton and anti-instanton contributions in Eq.~\eqref{eq:general-instanton-potential}, the magnitude of the susceptibility is given by the following expressions~\cite{Csaki:2023ziz}
\be
\label{eq:instanton-susceptibility-endpoints}
\chi\sim
2C_3\Lambda_G^{\,b_3 }
\prod_{i=1}^{r}m_i
\prod_{j=1}^k
\left(\frac{y_{j_1} y_{j_2} }{16\pi^2}\right)
\begin{cases}
\displaystyle
\left(\dfrac{2\pi}{\alpha_s(M_{\rm UV})}\right)^6
\dfrac{M_{\rm UV}^{-\Delta}}{-\Delta},
& \Delta<0 \quad (\text{UV dominated})\,, \\[3mm]
\displaystyle
\left(\dfrac{2\pi}{\alpha_s(\Lambda)}\right)^6
\ln\!\left(\dfrac{M_{\rm UV}}{\Lambda}\right),
& \Delta=0 \quad (\text{logarithmic}) \,, \\[3mm]
\displaystyle
\left(\dfrac{2\pi}{\alpha_s(\Lambda)}\right)^6
\dfrac{\Lambda^{-\Delta}}{\Delta},
& \Delta>0 \quad (\text{IR dominated})\,.
\end{cases}
\ee

As evident from the above Eq.~\eqref{eq:instanton-susceptibility-endpoints}, the UV-dominated contributions are sensitive to the highest scale over which the theory is extrapolated. 
We take this scale to be $M_{\rm UV}=M_{\rm Pl}$ and retain an instanton estimate only when $\alpha_s(\mu)<1$ throughout $\Lambda\leq\mu\leq M_{\rm Pl}$. 
Additionally, the involved degrees of freedom (VQs, scalars, SM fermions) are also charged under ${\rm SU}(2)_W$ and ${\rm U}(1)_Y$, and we impose the same condition of $\alpha_i(\mu)<1$ for all three SM gauge couplings~\footnote{We use the ${\rm SU}(5)$ GUT-normalized hypercharge coupling of $\alpha_1=(5/3)\alpha_Y$.}.

 The couplings that saturate the fermion zero modes can, in general, introduce a relative phase. 
 We assume that this phase is aligned with the QCD phase, so that the small-instanton and QCD potentials are minimized at the same value of $\bar\theta+a/f_a$. 
 Their susceptibilities then add constructively without displacing the CP-conserving minimum. 
Adding the instanton amplitude with the anti-instanton amplitude produces the following real contribution to the axion potential~\cite{tHooft:1976snw,Csaki:2023ziz}
\be
\label{eq:instanton-axion-potential}
V(a)=
-\chi\cos\!\left(\bar\theta+\frac{a}{f_a}\right)\,.
\ee
The coefficient $\chi$ is the topological susceptibility and fixes the axion mass through the relation of~\cite{GrillidiCortona:2015jxo,Gorghetto:2018ocs}
\be
\label{eq:susceptibility-and-axion-mass}
\chi=
\left.
\frac{\partial^2E(\theta)}
{\partial\theta^2}
\right|_{\theta=0}\,,
\quad
m_a^2=\frac{\chi}{f_a^2}\,.
\ee
 $E(\theta)$ denotes the vacuum-energy density as a function of the QCD vacuum angle $\theta$. 
 In the presence of the axion, $\theta$ angle is promoted to the effective combination $\theta=\bar\theta+a/f_a$, and $\theta=0$ denotes the CP-conserving minimum about which the curvature is evaluated. Eqs.~\eqref{eq:mass-insertion-factor}--\eqref{eq:nda-loop-factor} determine the factors associated with the zero-mode closures. Eqs.~\eqref{eq:instanton-size-power}--\eqref{eq:instanton-susceptibility-endpoints} then determine the dominant endpoint of the instanton-size integral and the resulting susceptibility. 
Finally, Eqs.~\eqref{eq:instanton-axion-potential} and \eqref{eq:susceptibility-and-axion-mass} translate this susceptibility into the axion potential and mass.

The different VQ representations listed in Tab.~\ref{tab:vq-copy-number-bounds} are taken from the Refs.~\cite{DiLuzio:2016sbl,DiLuzio:2017pfr,Plakkot:2021xyx,Cheek:2023fht,DiLuzio:2024xnt}. These studies incorporate the effects of VQ decays on the early cosmological history, including the possibility of an early matter-dominated era, and allow viable models whose leading decay operators have mass dimension as high as nine. We focus on VQs that satisfying the imposed perturbativity criterion and for which the small-instanton contribution exhibits appreciable UV sensitivity.
For each VQ representation, we construct the corresponding 't~Hooft vertex, identify the available mass and scalar-Yukawa closures, use their $\rho$ dependence to classify the instanton-size integral as UV dominated or IR dominated, and evaluate the resulting contribution to the axion susceptibility.

\section{UV sensitivity across KSVZ spectra}
\label{sec:vq-models}

We extend the SM by a Dirac VQ of $Q=Q_L+Q_R\sim(R_3\,,R_2,Y)$ and a complex SM-singlet scalar $\Phi$. 
$R_3$ and $R_2$, respectively, denote the ${\rm SU}(3)_c$ and the ${\rm SU}(2)_W$ representations, while $Y$ is the hypercharge. 
The relevant Lagrangian is
\be
\label{eq:vq-pq-lagrangian}
{\cal L}\supset {\cal L}_{\rm SM} +
|\partial_\mu\Phi|^2
+\overline Q i\gamma^\mu D_\mu Q
-\left[y_Q\Phi\,\overline Q_LQ_R\hc\right]
-\lambda_\Phi\left(
|\Phi|^2-\frac{v^2}{2}
\right)^2 \,,
\ee
with ${\cal L}_{\rm SM}$ representing the SM Lagrangian. 
The Lagrangian in Eq.~\eqref{eq:vq-pq-lagrangian} possesses a global ${\rm U}(1)_{\rm PQ}$ symmetry, under which the SM fermions are taken to be neutral. 
Denoting the PQ charges by $X$ and choosing the normalization $X(\Phi)=1$, the new fields transform as 
\be
\label{eq:pq-charge-difference}
\Phi\longrightarrow e^{i\alpha}\Phi,
\quad
Q_L\longrightarrow e^{iX_L\alpha}Q_L,
\quad
Q_R\longrightarrow e^{iX_R\alpha}Q_R,
\quad
\Delta X\equiv X_L-X_R=1.
\ee
 When $\Phi$ acquires a vacuum expectation value, the PQ symmetry is spontaneously broken. 
 The scalar field $\Phi$ is parametrized as 
\be
\label{eq:pq-field-parametrisation}
\Phi=
\frac{v+\sigma}{\sqrt{2}}\,
\exp\!\left(\frac{ia}{v}\right),
\quad
a\longrightarrow a+v\alpha,
\quad
v=N_{\rm DW}f_a \,.
\ee
In the above expression, $N_{\rm DW}$ denotes the domain-wall number, whose relation to the color anomaly is given in Eq.~\eqref{eq:vq-anomaly-coefficients}. 
The axion field $a(x)$ corresponds to the angular mode of $\Phi$, while $\sigma(x)$ is its radial excitation. 
Expanding about this vacuum gives the masses of the radial mode scalar field and the VQ as
\be
\label{eq:radial-and-vq-masses}
m_\sigma=\sqrt{2\lambda_\Phi}\,v,
\quad
M_Q=\frac{y_Qv}{\sqrt{2}} \,.
\ee
The color and electromagnetic anomaly coefficients can be computed using the following expressions;
\be
\label{eq:vq-anomaly-coefficients}
N=\Delta X\,T (R_3) {\rm dim}(R_2)\,,
\quad
E=\Delta X\, {\rm dim}(R_3)
\sum_{m=-j}^{j}(Y+m)^2\,,
\quad
N_{\rm DW}=2|N|,
\ee
where ${\rm dim}(R_2)=2j+1$ for an isospin-$j$ representation, and $m$ runs over its isospin components. The QCD anomaly explicitly breaks the continuous PQ symmetry. 
In particular, under the axion shift in Eq.~\eqref{eq:pq-field-parametrisation}, the one-instanton phase transforms as
\be
\label{eq:pq-transformation-instanton-phase}
\exp\!\left[i\left(\bar\theta+\frac{a}{f_a}\right)\right]
\longrightarrow
e^{iN_{\rm DW}\alpha}
\exp\!\left[i\left(\bar\theta+\frac{a}{f_a}\right)\right].
\ee
QCD instantons therefore preserve only the discrete subgroup of $\mathbb Z_{N_{\rm DW}}$ and generate the periodic axion potential. 
A QCD instanton produces zero modes for both chiral components of the Dirac VQ. For $\Delta X=1$, these modes form $2 T (R_3){\rm dim}(R_2)=2|N|$ conjugate pairs. 
Since each scalar-Yukawa loop saturates two zero-mode pairs and each mass insertion saturates one, complete saturation requires that
\be
\label{eq:vq-zero-mode-number}
2|N|=2T (R_3) {\rm dim}(R_2)=2k+r \,,
\ee
where $k$ counts scalar-Yukawa loops, and $r$ counts mass insertions. 
The color Dynkin indices $T(R_3)$ and the weak multiplicities ${\rm dim}(R_2)$ entering Eq.~\eqref{eq:vq-zero-mode-number} for different VQs of Tab.~\ref{tab:vq-copy-number-bounds} are collected in Appendix~\ref{sec:appendix}.

To evaluate the instanton-size integral in Eq.~\eqref{eq:general-instanton-potential}, one must specify the lower matching scale $\Lambda$, at which the active particle content changes. 
We identify this scale with the radial-mode mass~$(m_\sigma)$. 
For each VQ representation listed in Tab.~\ref{tab:vq-copy-number-bounds}, we collect the positive one-loop matter factors $\delta b_i$, the one- and two-loop copy-number bounds, the anomaly coefficients $E$ and $N$, and their ratio $E/N$. 
The derivation of the copy-number bounds is described below.

For the perturbativity analysis, we determine how many identical copies of each representation can remain perturbative up to the UV scale $M_{\rm UV}=M_{\rm Pl}=1.22\times10^{19}\,\GeV$. 
We evaluate the maximal copies at $\Lambda=5\times10^{11}\,\GeV$. 
Using the $\mathrm{SU}(5)$ GUT normalization, and defining $\alpha_i=g_i^2/(4\pi)$, the one-loop running of the gauge couplings is
\be
\label{eq:one-loop-alpha-running}
\frac{\mathrm d\alpha_i^{-1}}{\mathrm d\ln\mu}
=
\frac{b_i}{2\pi}\,,
\quad
b_i=b_i^{\rm SM}-n_Q\delta b_i\,,
\quad
b_i^{\rm SM}
=
\left(-\frac{41}{10},\frac{19}{6},7\right)\,,
\ee
where $\mu$ is the renormalization scale, $n_Q$ is the number of identical Dirac VQ copies, and $\delta b_i>0$ denotes the magnitude of the matter contribution from the VQs. 
Assuming a common VQ threshold at some $\Lambda$, integration up to the UV scale gives
\be
\label{eq:integrated-one-loop-alpha}
\alpha_i^{-1}(M_{\rm UV})
=
\alpha_i^{-1}(\Lambda)
+
\frac{b_i^{\rm SM}-n_Q\delta b_i}{2\pi}
\ln\!\left(\frac{M_{\rm UV}}{\Lambda}\right) \,.
\ee
We impose that $\alpha_i(M_{\rm Pl})<1$, equivalently $\alpha_i^{-1}(M_{\rm Pl})>1$. 
Each VQ lowers $b_i$ by the positive amount $\delta b_i$. 
Sufficiently many copies can make $b_i<0$, after which the coupling grows toward the UV region. 
The continuous boundary is obtained by saturating the perturbativity condition of $\alpha_i^{-1}(M_{\rm UV})=1$. 
Solving for $n_Q$ gives
\be
\label{eq:continuous-copy-budget}
n_{Q,i}^{\rm cont}
=
\frac{
\displaystyle
\frac{2\pi[\alpha_i^{-1}(\Lambda)-1]}
{\ln(M_{\rm UV}/\Lambda)}
+b_i^{\rm SM}}
{\delta b_i} \,.
\ee
The allowed multiplicities of Vqs must satisfy $n_Q<n_{Q,i}^{\rm cont}$. 
By writing $\lceil x\rceil$ for the ceiling function and hence the largest allowed integer is $n_{Q,i}^{\max}=\lceil n_{Q,i}^{\rm cont}\rceil-1$ which is equal to allowed maximum number of copies of respective VQs. 
The overall one-loop bound is $n_{Q,\max}^{(1)}=\min_i n_{Q,i}^{\max}$, that is, we take the minimum of $n_Q$ from three allowed gauge coupling bounds. 
If $\delta b_i=0$, that gauge factor imposes no one-loop bound. 
At $\Lambda=5\times10^{11}\,\GeV$, the inverse SM gauge couplings are
\be
\label{eq:matched-gauge-couplings}
\left(
\alpha_1^{-1},
\alpha_2^{-1},
\alpha_3^{-1}
\right)_{\Lambda}
=
(44.52,40.92,33.50)\,.
\ee
\begin{table*}[t!]
\centering
\normalsize
\setlength{\tabcolsep}{5pt}
\begin{tabular}{c c c c c c c c c c}
\hline\hline
Label & $VQ$ & $\delta b_1$ & $\delta b_2$ & $\delta b_3$
& $n_{Q,\max}^{(1)}$ & $n_{Q,\max}^{(2)}$
& $E$ & $N$ & $E/N$ \\
\hline
$R_1$ & $(3,1,-1/3)$  & $4/15$  & $0$    & $2/3$  & $28$ & $25$ & $1/3$  & $1/2$ & $2/3$ \\

$R_2$ & $(3,1, 2/3)$  & $16/15$ & $0$    & $2/3$  & $11$ & $11$ & $4/3$  & $1/2$ & $8/3$ \\

$R_3$ & $(3,2, 1/6)$  & $2/15$  & $2$    & $4/3$  & $8$  & $8$  & $5/3$  & $1$ & $5/3$ \\

$R_4$ & $(3,2,-5/6)$  & $10/3$  & $2$    & $4/3$  & $3$  & $3$  & $17/3$ & $1$ & $17/3$ \\

$R_5$ & $(3,2, 7/6)$  & $98/15$ & $2$    & $4/3$  & $1$  & $1$  & $29/3$ & $1$ & $29/3$ \\

$R_6$ & $(3,3,-1/3)$  & $4/5$   & $8$    & $2$    & $2$  & $2$  & $7$    & $3/2$ & $14/3$ \\

$R_7$ & $(3,3, 2/3)$  & $16/5$  & $8$    & $2$    & $2$  & $2$  & $10$   & $3/2$ & $20/3$ \\

$R_8$ & $(3,3,-4/3)$  & $64/5$  & $8$    & $2$    & $0$  & $0$  & $22$   & $3/2$ & $44/3$ \\

$R_9$ & $(6,1,-1/3)$  & $8/15$  & $0$    & $10/3$ & $5$  & $5$  & $2/3$  & $5/2$ & $4/15$ \\

$R_{10}$ & $(6,1, 2/3)$  & $32/15$ & $0$    & $10/3$ & $5$  & $4$  & $8/3$  & $5/2$ & $16/15$ \\

$R_{11}$ & $(6,2, 1/6)$  & $4/15$  & $4$    & $20/3$ & $2$  & $2$  & $10/3$ & $5$ & $2/3$ \\

$R_{12}$ & $(8,1,-1)$    & $32/5$  & $0$    & $4$    & $1$  & $1$  & $8$    & $3$ & $8/3$ \\

$R_{13}$ & $(8,2,-1/2)$  & $16/5$  & $16/3$ & $8$    & $2$  & $2$  & $8$    & $6$ & $4/3$ \\

$R_{14}$ &$(10,1,-1)$ & $8$ & $0$ & $10$ & $1$ & $1$ & $10$ & $15/2$ &$4/3$ \\

$R_{15}$ & $(10,2,-1/2)$ & $4$ & $20/3$ & $20$ & $0$ & $0$ & $10$ & $15$ & $2/3$ \\

$R_{16}$ & $(15,1,-1/3)$ & $4/3$   & $0$    & $40/3$ & $1$  & $1$  & $5/3$  & $10$ & $1/6$ \\

$R_{17}$ & $(15,1, 2/3)$ & $16/3$  & $0$    & $40/3$ & $1$  & $1$  & $20/3$ & $10$ & $2/3$ \\
\hline\hline
\end{tabular}
\caption{Gauge contributions, copy-number bounds and anomaly coefficients
for the different VQ candidates at $\Lambda=5\times10^{11}\,\GeV$.
The two-loop values are obtained by using $y_Q=1$ and the perturbativity condition of $\alpha_i(M_{\rm Pl})<1$. 
The entries of $E$ and $N$ are obtained by using $\Delta X=1$.
}
\label{tab:vq-copy-number-bounds}
\end{table*}

The contributions to $\delta b_i$ are listed in Tab.~\ref{tab:vq-copy-number-bounds}, and the group-theoretical expressions used to calculate them  are given in Appendix~\ref{sec:appendix}. 
The one-loop bounds obtained from Eq.~\eqref{eq:continuous-copy-budget} by using Eq.~\eqref{eq:matched-gauge-couplings} are provided in Tab.~\ref{tab:vq-copy-number-bounds} in the column $n_Q^{(1)}$. 
We repeat the perturbativity test at two loops as well, where the gauge beta functions contain mixed-gauge and Yukawa contributions. 
The two-loop RGEs were generated using PyR@TE~3~\cite{Sartore:2020gou}. 
For the two-loop entries in Tab.~\ref{tab:vq-copy-number-bounds}, we take $y_Q=1$, and the allowed bounds are relegated in the column $n_Q^{(2)}$.

For $y_Q=1$, $\Lambda=5\times10^{11}\,\GeV$ and
$\alpha_i(M_{\rm Pl})<1$, the two-loop evolution changes the maximum
number of copies only for $R_1$ and $R_{10}$, from $28$ to $25$ and from
$5$ to $4$, respectively. The larger change for $R_1$ follows from the
cumulative contribution of the many color triplets allowed at one loop.
For the other representations with a completed two-loop entry in
Tab.~\ref{tab:vq-copy-number-bounds}, the two-loop correction does not
change allowed copies at one loop.

Two representations fail the perturbativity requirement already for one
copy. The large hypercharge contribution of $R_8$ makes
$\alpha_1(M_{\rm Pl})>1$, whereas the contribution
$\delta b_3=20$ of $R_{15}$ makes $\alpha_3(M_{\rm Pl})>1$.\footnote{For
$R_8$, the one-loop evolution gives
$\left(\alpha_1^{-1},\alpha_2^{-1},\alpha_3^{-1}\right)_{M_{\rm Pl}}
=(-1.23,27.84,47.04)$. The hypercharge coupling reaches
$\alpha_1=1$ at $5.32\times10^{18}\,\GeV$ and develops a Landau pole at
$7.72\times10^{18}\,\GeV$. For
$R_{15}=(10,2,-1/2)$, one instead obtains
$\left(\alpha_1^{-1},\alpha_2^{-1},\alpha_3^{-1}\right)_{M_{\rm Pl}}
=(22.59,31.45,-1.69)$. In this case the ${\rm SU}(3)_c$ coupling reaches
$\alpha_3=1$ at $3.32\times10^{18}\,\GeV$ and develops a Landau pole at
$5.38\times10^{18}\,\GeV$. These two cases can be incorporated by increasing the matching scale~($\Lambda$) or lowering the UV scale~(M$_{\rm Pl}$).}
For a single VQ, Eqs.~\eqref{eq:continuous-copy-budget} and
\eqref{eq:matched-gauge-couplings} constrains
$\delta b_3\leq19$ which leads to an upper bound on the ${\rm SU} (3)$-Dynkin index as,
\be {\label{eq:SU3-dynkin-bound}}
T(R_3)<\frac{57}{4\,{\rm dim}(R_2)}\,.
\ee
For an ${\rm SU}(2)_W$ singlet this becomes $T(R_3)\leq57/4$, while for a
non-singlet weak representation makes the bound stronger. 
Tab.~\ref{tab:su3-group-invariants} shows the Dynkin indices of various ${\rm SU}(3)$ representations and it is evident that VQs with ${\rm dim}(R_3) > 15$ violates the inequality given in Eq.~\eqref{eq:SU3-dynkin-bound} and hence are not considered subsequently.

The matching-scale dependence is logarithmic, so changing $\Lambda$ by an order of magnitude shifts the largest integer bounds by only a few copies without altering them by order of magnitude. 
The one-loop bounds are independent of $y_Q$. 
In the two-loop running, varying an order-one $y_Q$ changes the largest bounds by at most one. The two-loop running is used only to determine which spectra remain perturbative up to $M_{\rm Pl}$. 
The instanton calculation below continues to use the one-loop coefficient $b_3$.
Throughout this article, $R_i$ denotes a single VQ representation listed in Tab.~\ref{tab:vq-copy-number-bounds}, whereas $R_i^{(n)}$ denotes a spectrum containing $n$ identical copies of that representation.

\subsection{Single VQ}
\label{subsec:one-vq}

\begin{figure*}[t!]
\centering
\includegraphics[width=0.30\textwidth]{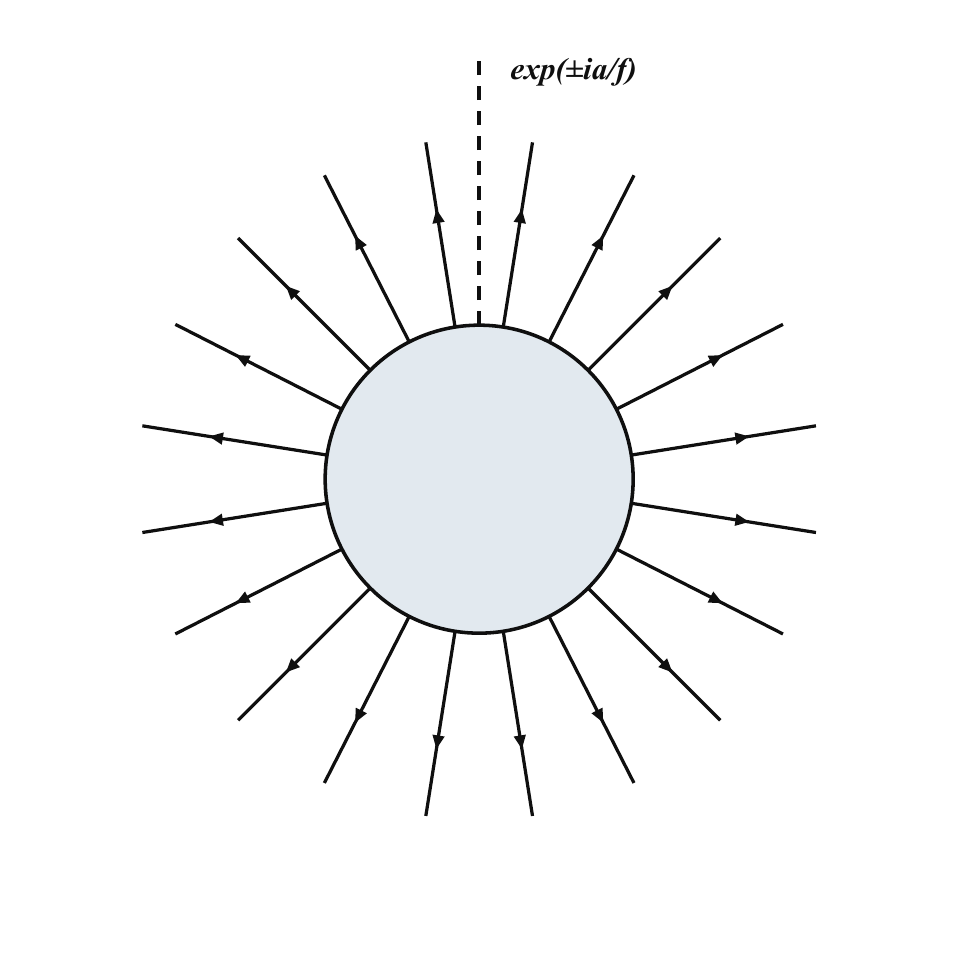}
\includegraphics[width=0.30\textwidth]{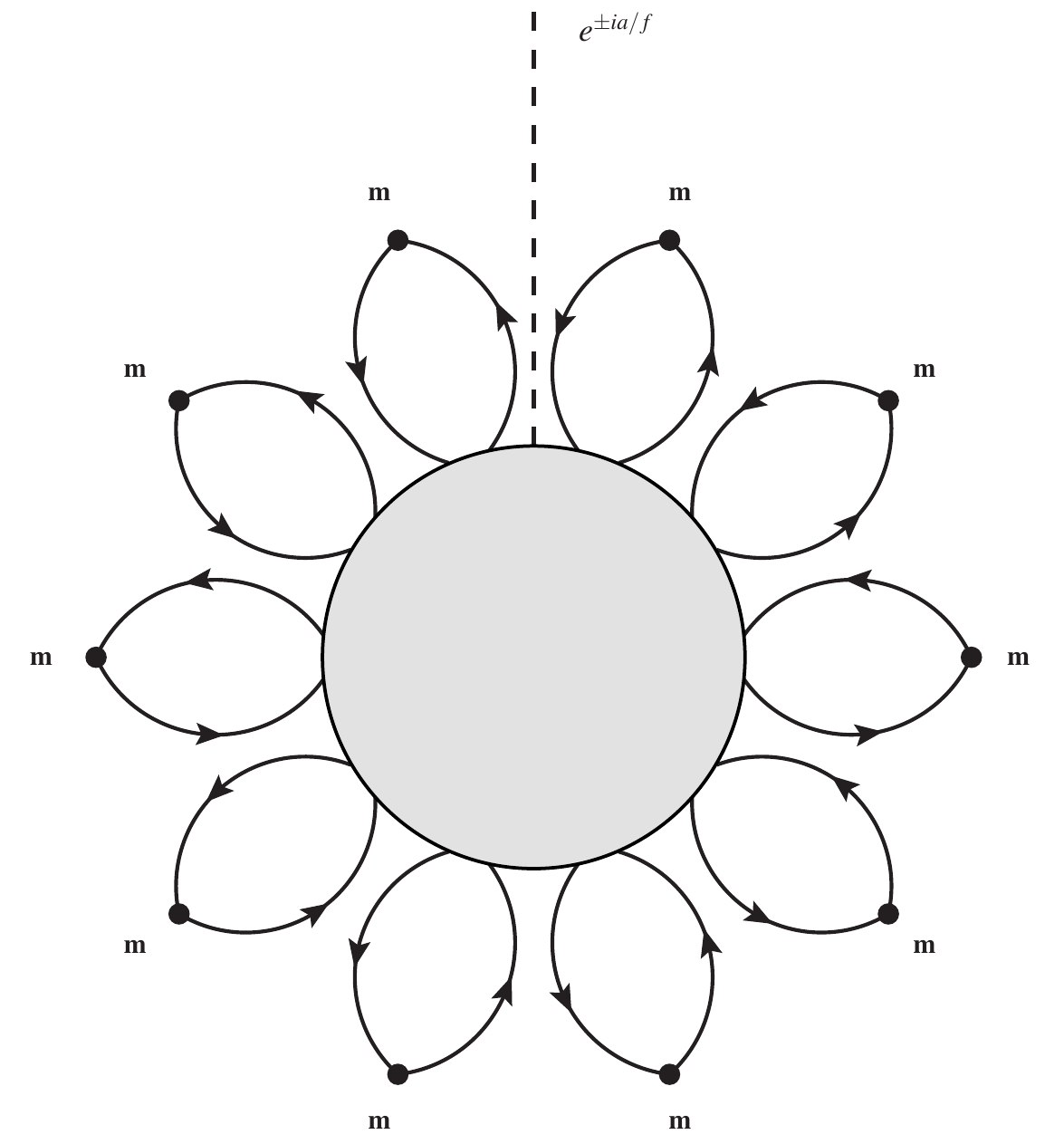}
\includegraphics[width=0.30\textwidth]{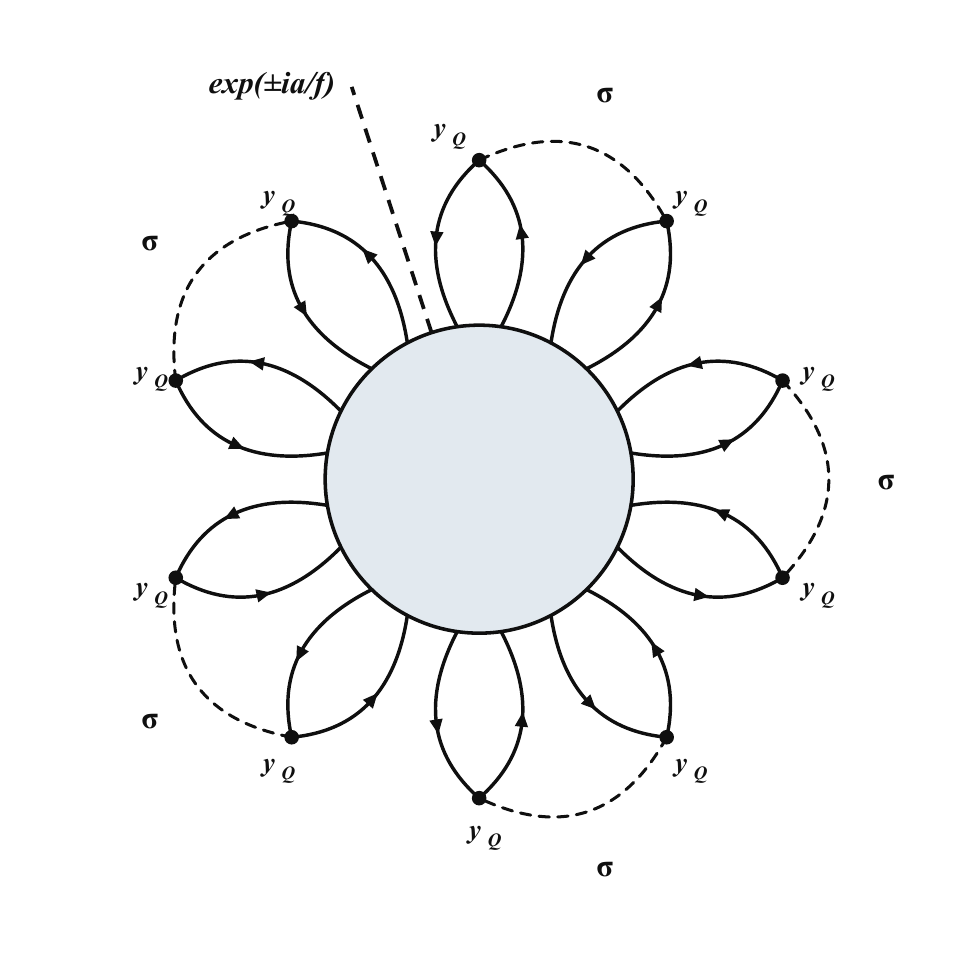}
\caption{The 't~Hooft vertex of ${\rm SU}(3)_c$ $15$-plet (left panel), maximally mass-closed contraction with $(k,r)=(0,20)$ (center panel), and Yukawa-closed contraction with $(k,r)=(10,0)$ (right panel). }
\label{fig:vq15-zero-mode-closures}
\end{figure*}

We begin the analysis by considering one VQ at a time and explicitly show the computation for the VQ that has the largest number of zero-mode pairs among the different VQs provided in Tab.~\ref{tab:vq-copy-number-bounds}. 
For the $15$-dimensional color and weak singlet representation~$(R_{16,\,17})$, the relevant group theoretical quantities are
\be
\label{eq:vq15-group-data}
T(15)=10\,,
\quad
2N=2T (15)=20\,,
\quad
b_3=7-\frac{2}{3}(2N)=-\frac{19}{3}\,,
\quad
C_3(15,1)=0.51\,.
\ee
These quantities are common to $Q=(15,1,-1/3)$ and $Q=(15,1,2/3)$, while their different hypercharges affect only the electroweak running. 
Fig.~\ref{fig:vq15-zero-mode-closures} shows the 't Hooft vertex for three cases of: no saturation~(left panel), the maximal mass insertions with $(k,r)=(0, 20)$ (center panel), and the maximally Yukawa-closed contraction with $(k,r)=(10,0)$ (right panel).

As is evident from Eq.~\eqref{eq:vq-zero-mode-number}, $2N$ counts the zero-mode pairs available for saturation. 
Let $k$ scalar-Yukawa loops close $2k$ such pairs. 
These scalar loops are due to the integration of radial field $\sigma$. 
The remaining $r=20-2k$ pairs are closed by VQ mass insertions. 
The resulting contractions, with $k=0,\ldots,10$, are separated into IR- and UV-dominated terms in Tab.~\ref{tab:one-vq-kr-pairs}. 



The endpoint exponent for this ${\rm SU}(3)_c$ $15-plet$ is given by
\be
\label{eq:vq15-endpoint-exponent}
\Delta
=
b_3 +r-4
=
\frac{29}{3}-2k\,,
\quad
k=0,\ldots,10\,,
\ee
according to Eq.~\eqref{eq:instanton-endpoint-exponent}.
The contractions with $k\leq4$ are IR dominated, whereas those with $k\geq5$ are UV dominated. 
Six SM quark flavors supply six additional zero-mode pairs, which we close through Higgs Yukawa couplings; these are in addition to the VQ zero modes. 
For the numerical estimates, SM Yukawa couplings are inferred from the quark masses ~\cite{ParticleDataGroup:2024cfk} and these masses are provided at different renormalization scale and we have not considered their RG effect.
We define
\be
\label{eq:sm-zero-mode-factor}
Y_{\rm SM}
=
\frac{y_u y_d y_c y_s y_t y_b}{(16\pi^2)^3}=8.24\times10^{-24}\,.
\ee
The matching of the threshold with the one-loop QCD running at $\Lambda$ gives
\be
\label{eq:qcd-threshold-matching}
\Lambda_G^{\,b_3 }
=
\Lambda_{\rm QCD}^{\,7}\Lambda^{\,b_3 -7} \,. 
\ee
For $R_{16,17}$, the IR- and UV-dominated contributions to the instanton induced susceptibility are given by
\beqa
\label{eq:vq15-ir-uv-contributions}
\chi_{(15,1,-1/3)}^{\rm IR}
\eq\chi_{(15,1,2/3)}^{\rm IR}\nl
\eq
2C_3(15,1)
\left(\frac{2\pi}{\alpha_s(\Lambda)}\right)^6
Y_{\rm SM}\Lambda_{\rm QCD}^{4}
\left(\frac{\Lambda_{\rm QCD}}{\Lambda}\right)^3 \times \nl
& & \Bigg[
\frac{3}{29}\left(\frac{M_Q}{\Lambda}\right)^{20}
+\frac{3}{23}\left(\frac{y_Q^2}{16\pi^2}\right)
\left(\frac{M_Q}{\Lambda}\right)^{18}
+\frac{3}{17}\left(\frac{y_Q^2}{16\pi^2}\right)^2
\left(\frac{M_Q}{\Lambda}\right)^{16}
\nl
\ad\frac{3}{11}\left(\frac{y_Q^2}{16\pi^2}\right)^3
\left(\frac{M_Q}{\Lambda}\right)^{14}
+\frac{3}{5}\left(\frac{y_Q^2}{16\pi^2}\right)^4
\left(\frac{M_Q}{\Lambda}\right)^{12}\Bigg] \,,
\nl
\chi_{(15,1,-1/3)}^{\rm UV}
\eq\chi_{(15,1,2/3)}^{\rm UV}\nl
\eq
2C_3(15,1)
\left(\frac{2\pi}{\alpha_s(M_{\rm UV})}\right)^6
Y_{\rm SM}\Lambda_{\rm QCD}^{4}
\left(\frac{\Lambda_{\rm QCD}}{\Lambda}\right)^3\,\times\nl
& &\Bigg[
3\left(\frac{y_Q^2}{16\pi^2}\right)^5
\left(\frac{M_Q}{\Lambda}\right)^{10}
\left(\frac{M_{\rm UV}}{\Lambda}\right)^{1/3}
+
\frac{3}{7}\left(\frac{y_Q^2}{16\pi^2}\right)^6
\left(\frac{M_Q}{\Lambda}\right)^8
\left(\frac{M_{\rm UV}}{\Lambda}\right)^{7/3}
\nl
\ad\frac{3}{13}\left(\frac{y_Q^2}{16\pi^2}\right)^7
\left(\frac{M_Q}{\Lambda}\right)^6
\left(\frac{M_{\rm UV}}{\Lambda}\right)^{13/3}
+\frac{3}{19}\left(\frac{y_Q^2}{16\pi^2}\right)^8
\left(\frac{M_Q}{\Lambda}\right)^4
\left(\frac{M_{\rm UV}}{\Lambda}\right)^{19/3}
\nl
\ad\frac{3}{25}\left(\frac{y_Q^2}{16\pi^2}\right)^9
\left(\frac{M_Q}{\Lambda}\right)^2
\left(\frac{M_{\rm UV}}{\Lambda}\right)^{25/3}
+\frac{3}{31}\left(\frac{y_Q^2}{16\pi^2}\right)^{10}
\left(\frac{M_{\rm UV}}{\Lambda}\right)^{31/3}\Bigg] \,. 
\eeqa
For $M_{\rm UV}\gg\Lambda$, the fully Yukawa-closed term with $(k,r)=(10,0)$ has the strongest enhancement, $(M_{\rm UV}/\Lambda)^{31/3}$, and dominates the UV contribution. 
Table~\ref{tab:one-vq-kr-pairs} provides the remaining VQs, and fixes their
$SU(3)_c$ zero-mode counting. The corresponding
susceptibilities for remaining VQs are given below.


\begin{table*}[t!]
\centering
\normalsize
\setlength{\tabcolsep}{4.5pt}
\begin{tabular}{l c c l l}
\hline\hline
~~~~~VQs~~~~~ & ~~~$2N$~~~ & ~~~$b_3$~~~ & IR $(k,r)$ & UV $(k,r)$\\
\hline
$(3,1,-1/3)$, $(3,1,2/3)$
& $1$ & $19/3$ & $(0,1)$ & none\\
\begin{tabular}[c]{@{}l@{}}$(3,2,1/6)$, $(3,2,-5/6)$\\$(3,2,7/6)$\end{tabular}
& $2$ & $17/3$ & $(0,2),(1,0)$ & none\\
$(3,3,-1/3)$, $(3,3,2/3)$
& $3$ & $5$ & $(0,3),(1,1)$ & none\\
$(6,1,-1/3)$, $(6,1,2/3)$
& $5$ & $11/3$ & $(0,5),(1,3),(2,1)$ & none\\
$(6,2,1/6)$
& $10$ & $1/3$
& \begin{tabular}[c]{@{}l@{}}$(0,10),(1,8)$\\$(2,6),(3,4)$\end{tabular}
& $(4,2),(5,0)$\\
$(8,1,-1)$
& $6$ & $3$ & $(0,6),(1,4),(2,2)$ & $(3,0)$\\
$(8,2,-1/2)$
& $12$ & $-1$
& \begin{tabular}[c]{@{}l@{}}$(0,12),(1,10)$\\$(2,8),(3,6)$\end{tabular}
& $(4,4),(5,2),(6,0)$\\
$(10,1,-1)$& $15$ & $-3$& \begin{tabular}[c]{@{}l@{}}$(0,15),(1,13)$\\$(2,11),(3,9),(4,7)$\end{tabular}& $(5,5),(6,3),(7,1)$\\
$(15,1,-1/3)$, $(15,1,2/3)$
& $20$ & $-19/3$
& \begin{tabular}[c]{@{}l@{}}$(0,20),(1,18),(2,16)$\\$(3,14),(4,12)$\end{tabular}
& \begin{tabular}[c]{@{}l@{}}$(5,10),(6,8),(7,6)$\\$(8,4),(9,2),(10,0)$\end{tabular}\\
\hline\hline
\end{tabular}
\caption{The table lists the number of zero-mode pairs, $b_3$, and the allowed pairs $(k,r)$ satisfying Eq.~\eqref{eq:zero-mode-saturation} for single VQs. Representations with identical ${\rm SU}(3)_c$ and ${\rm SU}(2)_W$ charged are grouped together.}
\label{tab:one-vq-kr-pairs}
\end{table*}
For the color-triplet weak singlets of $Q=(3,1,-1/3)$ and $Q=(3,1,2/3)$~$(R_{1,2})$, one obtains
\beqa
\label{eq:vq-triplet-singlet-ir-uv}
\chi_{(3,1,-1/3)}
=\chi_{(3,1,2/3)}\sim
2C_3(3,1)
\left(\frac{2\pi}{\alpha_s(\Lambda)}\right)^6
Y_{\rm SM}\Lambda_{\rm QCD}^{4}
\left(\frac{\Lambda_{\rm QCD}}{\Lambda}\right)^3
\frac{3}{10}\left(\frac{M_Q}{\Lambda}\right)\,.
\eeqa
Three color-triplet weak doublets~$(R_{3,4,5})$ have a common instanton induced IR sensitive susceptibility of
\beqa
\label{eq:vq-triplet-doublet-ir-uv}
\chi_{(3,2,1/6)}
&=&\chi_{(3,2,-5/6)}
=\chi_{(3,2,7/6)}\nl
&\sim&
2C_3(3,2)
\left(\frac{2\pi}{\alpha_s(\Lambda)}\right)^6
Y_{\rm SM}\Lambda_{\rm QCD}^{4}
\left(\frac{\Lambda_{\rm QCD}}{\Lambda}\right)^3 \times \nl 
&& \left[\frac{3}{11}\left(\frac{M_Q}{\Lambda}\right)^2
+\frac{3}{5}\left(\frac{y_Q^2}{16\pi^2}\right)\right]\,. 
\eeqa

The color-triplet weak triplets of $Q=(3,3,-1/3)$ and $Q=(3,3,2/3)$~$(R_{6,7})$ also have a common instanton contribution of 
\beqa
\label{eq:vq-triplet-weak-triplet-ir-uv}
\chi_{(3,3,-1/3)}^{\rm IR}
\eq \chi_{(3,3,2/3)}^{\rm IR} \nl 
&\sim&
2C_3(3,3)
\left(\frac{2\pi}{\alpha_s(\Lambda)}\right)^6
Y_{\rm SM}\Lambda_{\rm QCD}^{4}
\left(\frac{\Lambda_{\rm QCD}}{\Lambda}\right)^3\times 
\left[\frac{1}{4}\left(\frac{M_Q}{\Lambda}\right)^3
+\frac{1}{2}\left(\frac{y_Q^2}{16\pi^2}\right)
\left(\frac{M_Q}{\Lambda}\right)\right] \nl \,. 
\eeqa
$R_{6,7}$ similarly exhibits only IR sensitive instanton induced susceptibility. For two color-sextet weak singlets of $Q=(6,1,-1/3)$ and $Q=(6,1,2/3)$~$(R_{9,10})$, the similar expression is
\beqa
\label{eq:vq-sextet-singlet-ir}
\chi_{(6,1,-1/3)}^{\rm IR}
\eq\chi_{(6,1,2/3)}^{\rm IR}\nl
&\sim&
2C_3(6,1)
\left(\frac{2\pi}{\alpha_s(\Lambda)}\right)^6
Y_{\rm SM}\Lambda_{\rm QCD}^{4}
\left(\frac{\Lambda_{\rm QCD}}{\Lambda}\right)^3
\Bigg[
\frac{3}{14}\left(\frac{M_Q}{\Lambda}\right)^5
+\frac{3}{8}\left(\frac{y_Q^2}{16\pi^2}\right)
\left(\frac{M_Q}{\Lambda}\right)^3
\nl
\ad\frac{3}{2}\left(\frac{y_Q^2}{16\pi^2}\right)^2
\left(\frac{M_Q}{\Lambda}\right)\Bigg] \,. 
\eeqa
The susceptibilities of different VQs $(R_{1-7,9,10})$ given by Eqs.~\eqref{eq:vq-triplet-singlet-ir-uv}--\eqref{eq:vq-sextet-singlet-ir} are therefore IR dominated. 
For the color-sextet weak doublet of $Q=(6,2,1/6)$~$(R_{11})$, instanton induces both UV and IR sensitive susceptibility, and is given by
\beqa
\label{eq:vq-sextet-doublet-ir-uv}
\chi_{(6,2,1/6)}^{\rm IR}
&\sim&
2C_3(6,2)
\left(\frac{2\pi}{\alpha_s(\Lambda)}\right)^6
Y_{\rm SM}\Lambda_{\rm QCD}^{4}
\left(\frac{\Lambda_{\rm QCD}}{\Lambda}\right)^3\times\nl
& &
\Bigg[
\frac{3}{19}\left(\frac{M_Q}{\Lambda}\right)^{10}
+\frac{3}{13}\left(\frac{y_Q^2}{16\pi^2}\right)
\left(\frac{M_Q}{\Lambda}\right)^8
+\frac{3}{7}\left(\frac{y_Q^2}{16\pi^2}\right)^2
\left(\frac{M_Q}{\Lambda}\right)^6\nl
\ad3\left(\frac{y_Q^2}{16\pi^2}\right)^3
\left(\frac{M_Q}{\Lambda}\right)^4\Bigg]\,, \nl
\chi_{(6,2,1/6)}^{\rm UV}
&\sim&
2C_3(6,2)
\left(\frac{2\pi}{\alpha_s(M_{\rm UV})}\right)^6
Y_{\rm SM}\Lambda_{\rm QCD}^{4}
\left(\frac{\Lambda_{\rm QCD}}{\Lambda}\right)^3\times\nl
& &
\Bigg[
\frac{3}{5}\left(\frac{y_Q^2}{16\pi^2}\right)^4
\left(\frac{M_Q}{\Lambda}\right)^2
\left(\frac{M_{\rm UV}}{\Lambda}\right)^{5/3}
+\frac{3}{11}\left(\frac{y_Q^2}{16\pi^2}\right)^5
\left(\frac{M_{\rm UV}}{\Lambda}\right)^{11/3}\Bigg]\,.
\eeqa

The UV and IR sensitive susceptibility due to instanton for the color-octet weak singlet of $Q=(8,1,-1)$~$(R_{12})$ is 
\beqa
\label{eq:vq-octet-singlet-ir-uv}
\chi_{(8,1,-1)}^{\rm IR}
&\sim&
2C_3(8,1)
\left(\frac{2\pi}{\alpha_s(\Lambda)}\right)^6
Y_{\rm SM}\Lambda_{\rm QCD}^{4}
\left(\frac{\Lambda_{\rm QCD}}{\Lambda}\right)^3\times\nl
& &
\Bigg[
\frac{1}{5}\left(\frac{M_Q}{\Lambda}\right)^6
+\frac{1}{3}\left(\frac{y_Q^2}{16\pi^2}\right)
\left(\frac{M_Q}{\Lambda}\right)^4
+\left(\frac{y_Q^2}{16\pi^2}\right)^2
\left(\frac{M_Q}{\Lambda}\right)^2\Bigg]\,,\nl
\chi_{(8,1,-1)}^{\rm UV}
&\sim&
2C_3(8,1)
\left(\frac{2\pi}{\alpha_s(M_{\rm UV})}\right)^6
Y_{\rm SM}\Lambda_{\rm QCD}^{4}
\left(\frac{\Lambda_{\rm QCD}}{\Lambda}\right)^3
\left(\frac{y_Q^2}{16\pi^2}\right)^3
\left(\frac{M_{\rm UV}}{\Lambda}\right)\,. 
\eeqa

For the color-octet weak doublet of $Q=(8,2,-1/2)$~$(R_{13})$, the corresponding terms are 
\beqa
\label{eq:vq-octet-doublet-ir-uv}
\chi_{(8,2,-1/2)}^{\rm IR}
&\sim&
2C_3(8,2)
\left(\frac{2\pi}{\alpha_s(\Lambda)}\right)^6
Y_{\rm SM}\Lambda_{\rm QCD}^{4}
\left(\frac{\Lambda_{\rm QCD}}{\Lambda}\right)^3\times\nl
& &
\Bigg[
\frac{1}{7}\left(\frac{M_Q}{\Lambda}\right)^{12}
+\frac{1}{5}\left(\frac{y_Q^2}{16\pi^2}\right)
\left(\frac{M_Q}{\Lambda}\right)^{10}\nl
\ad\frac{1}{3}\left(\frac{y_Q^2}{16\pi^2}\right)^2
\left(\frac{M_Q}{\Lambda}\right)^8
+\left(\frac{y_Q^2}{16\pi^2}\right)^3
\left(\frac{M_Q}{\Lambda}\right)^6\Bigg]\,,\nl
\chi_{(8,2,-1/2)}^{\rm UV}
&\sim&
2C_3(8,2)
\left(\frac{2\pi}{\alpha_s(M_{\rm UV})}\right)^6
Y_{\rm SM}\Lambda_{\rm QCD}^{4}
\left(\frac{\Lambda_{\rm QCD}}{\Lambda}\right)^3\times\nl
& &
\Bigg[
\left(\frac{y_Q^2}{16\pi^2}\right)^4
\left(\frac{M_Q}{\Lambda}\right)^4
\left(\frac{M_{\rm UV}}{\Lambda}\right)
+\frac{1}{3}\left(\frac{y_Q^2}{16\pi^2}\right)^5
\left(\frac{M_Q}{\Lambda}\right)^2
\left(\frac{M_{\rm UV}}{\Lambda}\right)^3\nl
\ad\frac{1}{5}\left(\frac{y_Q^2}{16\pi^2}\right)^6
\left(\frac{M_{\rm UV}}{\Lambda}\right)^5\Bigg]\,.
\eeqa

The color-decuplet weak singlet $Q=(10,1,-1)$~$(R_{14})$, also exhibits logarithmic contribution to susceptibility in addition to IR and UV and is given below
\beqa\label{eq:vq-decuplet-singlet-ir-log-uv}
\chi_{(10,1,-1)}^{\rm IR}&\sim&2C_3(10,1)\left(\frac{2\pi}{\alpha_s(\Lambda)}\right)^6Y_{\rm SM}\Lambda_{\rm QCD}^{4}\left(\frac{\Lambda_{\rm QCD}}{\Lambda}\right)^3\times\nl
&&\Bigg[\frac{1}{8}\left(\frac{M_Q}{\Lambda}\right)^{15}+\frac{1}{6}\left(\frac{y_Q^2}{16\pi^2}\right)\left(\frac{M_Q}{\Lambda}\right)^{13}\nl
\ad\frac{1}{4}\left(\frac{y_Q^2}{16\pi^2}\right)^2\left(\frac{M_Q}{\Lambda}\right)^{11}+\frac{1}{2}\left(\frac{y_Q^2}{16\pi^2}\right)^3\left(\frac{M_Q}{\Lambda}\right)^9\Bigg]\,,\nl
\chi_{(10,1,-1)}^{\log}&\sim&2C_3(10,1)\left(\frac{2\pi}{\alpha_s(\Lambda)}\right)^6Y_{\rm SM}\Lambda_{\rm QCD}^{4}\left(\frac{\Lambda_{\rm QCD}}{\Lambda}\right)^3
\left(\frac{y_Q^2}{16\pi^2}\right)^4\left(\frac{M_Q}{\Lambda}\right)^7\ln\left(\frac{M_{\rm UV}}{\Lambda}\right)\,,\nl
\chi_{(10,1,-1)}^{\rm UV}&\sim&2C_3(10,1)\left(\frac{2\pi}{\alpha_s(M_{\rm UV})}\right)^6Y_{\rm SM}\Lambda_{\rm QCD}^{4}\left(\frac{\Lambda_{\rm QCD}}{\Lambda}\right)^3\times\nl
&&\Bigg[\frac{1}{2}\left(\frac{y_Q^2}{16\pi^2}\right)^5\left(\frac{M_Q}{\Lambda}\right)^5\left(\frac{M_{\rm UV}}{\Lambda}\right)^2
+\frac{1}{4}\left(\frac{y_Q^2}{16\pi^2}\right)^6\left(\frac{M_Q}{\Lambda}\right)^3\left(\frac{M_{\rm UV}}{\Lambda}\right)^4\nl
\ad\frac{1}{6}\left(\frac{y_Q^2}{16\pi^2}\right)^7\left(\frac{M_Q}{\Lambda}\right)\left(\frac{M_{\rm UV}}{\Lambda}\right)^6\Bigg]\,.
\eeqa
For the decuplet, $(k,r)=(4,7)$ gives a logarithmic contribution, while $(k,r)=(7,1)$ gives the largest UV contribution scaling as $(M_{\rm UV}/\Lambda)^6$.

The color-triplet singlets ($R_{1,2}$), doublets ($R_{3-5}$), and triplets ($R_{6-8}$), together with the color-sextet singlets ($R_{9,10}$), admit only IR-dominated contractions. 
UV-sensitive terms occur for the color-sextet doublet ($R_{11}$), the color-octet singlet and doublet ($R_{12,13}$), the color-decuplet singlet ($R_{14}$), and the color-$15$ singlets ($R_{16,17}$).
For $M_{\rm UV}\gg\Lambda$ and perturbative $y_Q$, the contractions containing the largest number of scalar--Yukawa insertions are generally dominant. 
The strongest single-VQ sensitivity is obtained for the color-$15$ singlets ($R_{16,17}$) and scales as $(M_{\rm UV}/\Lambda)^{31/3}$. 
The determinant factors $C_3(R_3,R_2)$ entering these expressions are given in Eq.~\eqref{eq:vq-instanton-prefactor} of Appendix~\ref{sec:appendix}.

\subsection{Identical VQ copies}
\label{subsec:identical-vqs}

We next consider $n_Q$ identical copies of the same VQ representation and maximum multiplicities allowed for the same by gauge perturbativity are listed in Tab.~\ref{tab:vq-copy-number-bounds}. 
We take the VQ copies to have a common mass $M_Q$, Yukawa coupling $y_Q$, and PQ-anomaly sign. 
Finally, the octet doublet $R_{13}^{(2)}$ yields the UV-sensitive susceptibility 
Since a single VQ contains $2N$ conjugate zero-mode pairs in Eqs.~\eqref{eq:zero-mode-saturation} and \eqref{eq:instanton-endpoint-exponent}, $n_Q$ copies contain $2n_QN$ conjugate zero-mode pairs. The larger zero-mode multiplicity permits additional scalar--Yukawa contractions and can make the instanton-size integral UV dominated even when the corresponding single-VQ contribution is IR dominated.
The determinant factor for $n_Q$ identical copies can be written in terms of the single-copy factor as,
\be
\label{eq:identical-copy-prefactor}
C_3^{(n_Q)}(R_3\,,R_2)
=C_3(R_3\,,R_2)\exp\!\left[2(n_Q-1){\rm dim}(R_2)A(R_3)\right]\,.
\ee
The values of $A(R_3)$ and expression for the single-copy determinant factor $C_3(R_3,R_2)$ are provided in Appendix~\ref{sec:appendix}. In the expressions below we set $n_Q=n_{Q,\max}^{(2)}$ and retain only representations for which $n_{Q,\max}^{(2)}>1$ and the instanton-size integral contains UV-dominated terms.
 The two-loop bound permits $n_Q=25$ copies of $Q=(3,1,-1/3)$~$(R_1^{(25)})$, for which the UV sensitive susceptibility takes the form of
\beqa
\label{eq:identical-triplet-singlet-minus-uv}
\chi_{(3,1,-1/3)}^{\rm UV}
&\sim&
2C_3^{(25)}(3,1)\!\!\left(\frac{2\pi}{\alpha_s(M_{\rm UV})}\right)^6
Y_{\rm SM}\Lambda_{\rm QCD}^{4}
\left(\frac{\Lambda_{\rm QCD}}{\Lambda}\right)^3\Bigg[
\frac{3}{2}\left(\frac{y_Q^2}{16\pi^2}\right)^6
\left(\frac{M_Q}{\Lambda}\right)^{13}
\left(\frac{M_{\rm UV}}{\Lambda}\right)^{2/3}\nl
\ad\frac{3}{8}\left(\frac{y_Q^2}{16\pi^2}\right)^7
\left(\frac{M_Q}{\Lambda}\right)^{11}
\left(\frac{M_{\rm UV}}{\Lambda}\right)^{8/3}
+\frac{3}{14}\left(\frac{y_Q^2}{16\pi^2}\right)^8
\left(\frac{M_Q}{\Lambda}\right)^9
\left(\frac{M_{\rm UV}}{\Lambda}\right)^{14/3}\nl
\ad\frac{3}{20}\left(\frac{y_Q^2}{16\pi^2}\right)^9
\left(\frac{M_Q}{\Lambda}\right)^7
\left(\frac{M_{\rm UV}}{\Lambda}\right)^{20/3}
+\frac{3}{26}\left(\frac{y_Q^2}{16\pi^2}\right)^{10}
\left(\frac{M_Q}{\Lambda}\right)^5
\left(\frac{M_{\rm UV}}{\Lambda}\right)^{26/3}\nl
\ad\frac{3}{32}\left(\frac{y_Q^2}{16\pi^2}\right)^{11}
\left(\frac{M_Q}{\Lambda}\right)^3
\left(\frac{M_{\rm UV}}{\Lambda}\right)^{32/3}
+\frac{3}{38}\left(\frac{y_Q^2}{16\pi^2}\right)^{12}
\frac{M_Q}{\Lambda}
\left(\frac{M_{\rm UV}}{\Lambda}\right)^{38/3}\Bigg]\,.
\eeqa

For $R_2^{(11)}$, the UV-sensitive instanton contribution is
\beqa
\label{eq:identical-triplet-singlet-plus-uv}
\chi_{(3,1,2/3)}^{\rm UV}
&\sim&
2C_3^{(11)}(3,1)\left(\frac{2\pi}{\alpha_s(M_{\rm UV})}\right)^6
Y_{\rm SM}\Lambda_{\rm QCD}^{4}
\left(\frac{\Lambda_{\rm QCD}}{\Lambda}\right)^3\Bigg[
\frac{3}{4}\left(\frac{y_Q^2}{16\pi^2}\right)^4
\left(\frac{M_Q}{\Lambda}\right)^3
\left(\frac{M_{\rm UV}}{\Lambda}\right)^{4/3}\nl
\ad\frac{3}{10}\left(\frac{y_Q^2}{16\pi^2}\right)^5
\left(\frac{M_Q}{\Lambda}\right)
\left(\frac{M_{\rm UV}}{\Lambda}\right)^{10/3}\Bigg]\,.
\eeqa

The eight-copy triplet-doublet VQ $R_3^{(8)}$ gives
\beqa
\label{eq:identical-triplet-doublet-one-sixth-uv}
\chi_{(3,2,1/6)}^{\rm UV}
&\sim&
2C_3^{(8)}(3,2)\left(\frac{2\pi}{\alpha_s(M_{\rm UV})}\right)^6
Y_{\rm SM}\Lambda_{\rm QCD}^{4}
\left(\frac{\Lambda_{\rm QCD}}{\Lambda}\right)^3\Bigg[
\frac{3}{5}\left(\frac{y_Q^2}{16\pi^2}\right)^5
\left(\frac{M_Q}{\Lambda}\right)^6
\left(\frac{M_{\rm UV}}{\Lambda}\right)^{5/3}\nl
\ad\frac{3}{11}\left(\frac{y_Q^2}{16\pi^2}\right)^6
\left(\frac{M_Q}{\Lambda}\right)^4
\left(\frac{M_{\rm UV}}{\Lambda}\right)^{11/3}
+\frac{3}{17}\left(\frac{y_Q^2}{16\pi^2}\right)^7
\left(\frac{M_Q}{\Lambda}\right)^2
\left(\frac{M_{\rm UV}}{\Lambda}\right)^{17/3}\nl
\ad\frac{3}{23}\left(\frac{y_Q^2}{16\pi^2}\right)^8
\left(\frac{M_{\rm UV}}{\Lambda}\right)^{23/3}\Bigg]\,.
\eeqa

For $R_4^{(3)}$, one obtains the susceptibility as
\beqa
\label{eq:identical-triplet-doublet-minus-five-sixths-uv}
\chi_{(3,2,-5/6)}^{\rm UV}
&\sim&
2C_3^{(3)}(3,2)\left(\frac{2\pi}{\alpha_s(M_{\rm UV})}\right)^6
Y_{\rm SM}\Lambda_{\rm QCD}^{4}
\left(\frac{\Lambda_{\rm QCD}}{\Lambda}\right)^3
\left(\frac{y_Q^2}{16\pi^2}\right)^3
\left(\frac{M_{\rm UV}}{\Lambda}\right) \,.
\eeqa

The contribution of $R_6^{(2)}$ is
\beqa
\label{eq:identical-triplet-triplet-minus-uv}
\chi_{(3,3,-1/3)}^{\rm UV}
&\sim&
2C_3^{(2)}(3,3)\left(\frac{2\pi}{\alpha_s(M_{\rm UV})}\right)^6
Y_{\rm SM}\Lambda_{\rm QCD}^{4}
\left(\frac{\Lambda_{\rm QCD}}{\Lambda}\right)^3
\left(\frac{y_Q^2}{16\pi^2}\right)^3
\left(\frac{M_{\rm UV}}{\Lambda}\right) \,.
\eeqa

The case $R_7^{(2)}$ has the following behavior of instanton induced susceptibility,
\beqa
\label{eq:identical-triplet-triplet-plus-uv}
\chi_{(3,3,2/3)}^{\rm UV}
&\sim&
2C_3^{(2)}(3,3)\left(\frac{2\pi}{\alpha_s(M_{\rm UV})}\right)^6
Y_{\rm SM}\Lambda_{\rm QCD}^{4}
\left(\frac{\Lambda_{\rm QCD}}{\Lambda}\right)^3
\left(\frac{y_Q^2}{16\pi^2}\right)^3
\left(\frac{M_{\rm UV}}{\Lambda}\right)\,.
\eeqa

For the five-copy sextet configuration $R_9^{(5)}$, the UV-dominated contribution is
\beqa
\label{eq:identical-sextet-singlet-minus-uv}
\chi_{(6,1,-1/3)}^{\rm UV}
&\sim&
2C_3^{(5)}(6,1)\!\!\left(\frac{2\pi}{\alpha_s(M_{\rm UV})}\right)^6
Y_{\rm SM}\Lambda_{\rm QCD}^{4}
\left(\frac{\Lambda_{\rm QCD}}{\Lambda}\right)^3\Bigg[
\frac{3}{2}\left(\frac{y_Q^2}{16\pi^2}\right)^6
\left(\frac{M_Q}{\Lambda}\right)^{13}
\left(\frac{M_{\rm UV}}{\Lambda}\right)^{2/3}\nl
\ad\frac{3}{8}\left(\frac{y_Q^2}{16\pi^2}\right)^7
\left(\frac{M_Q}{\Lambda}\right)^{11}
\left(\frac{M_{\rm UV}}{\Lambda}\right)^{8/3}
+\frac{3}{14}\left(\frac{y_Q^2}{16\pi^2}\right)^8
\left(\frac{M_Q}{\Lambda}\right)^9
\left(\frac{M_{\rm UV}}{\Lambda}\right)^{14/3}\nl
\ad\frac{3}{20}\left(\frac{y_Q^2}{16\pi^2}\right)^9
\left(\frac{M_Q}{\Lambda}\right)^7
\left(\frac{M_{\rm UV}}{\Lambda}\right)^{20/3}
+\frac{3}{26}\left(\frac{y_Q^2}{16\pi^2}\right)^{10}
\left(\frac{M_Q}{\Lambda}\right)^5
\left(\frac{M_{\rm UV}}{\Lambda}\right)^{26/3}\nl
\ad\frac{3}{32}\left(\frac{y_Q^2}{16\pi^2}\right)^{11}
\left(\frac{M_Q}{\Lambda}\right)^3
\left(\frac{M_{\rm UV}}{\Lambda}\right)^{32/3}
+\frac{3}{38}\left(\frac{y_Q^2}{16\pi^2}\right)^{12}
\frac{M_Q}{\Lambda}
\left(\frac{M_{\rm UV}}{\Lambda}\right)^{38/3}\Bigg] \,.
\eeqa

The sextet singlet, $R_{10}^{(4)}$, gives instanton induced susceptibility as
\beqa
\label{eq:identical-sextet-singlet-plus-uv}
\chi_{(6,1,2/3)}^{\rm UV}
&\sim&
2C_3^{(4)}(6,1)\left(\frac{2\pi}{\alpha_s(M_{\rm UV})}\right)^6
Y_{\rm SM}\Lambda_{\rm QCD}^{4}
\left(\frac{\Lambda_{\rm QCD}}{\Lambda}\right)^3\times\nl
& &\Bigg[
3\left(\frac{y_Q^2}{16\pi^2}\right)^5
\left(\frac{M_Q}{\Lambda}\right)^{10}
\left(\frac{M_{\rm UV}}{\Lambda}\right)^{1/3}
+\frac{3}{7}\left(\frac{y_Q^2}{16\pi^2}\right)^6
\left(\frac{M_Q}{\Lambda}\right)^8
\left(\frac{M_{\rm UV}}{\Lambda}\right)^{7/3}\nl
\ad\frac{3}{13}\left(\frac{y_Q^2}{16\pi^2}\right)^7
\left(\frac{M_Q}{\Lambda}\right)^6
\left(\frac{M_{\rm UV}}{\Lambda}\right)^{13/3}
+\frac{3}{19}\left(\frac{y_Q^2}{16\pi^2}\right)^8
\left(\frac{M_Q}{\Lambda}\right)^4
\left(\frac{M_{\rm UV}}{\Lambda}\right)^{19/3}\nl
\ad\frac{3}{25}\left(\frac{y_Q^2}{16\pi^2}\right)^9
\left(\frac{M_Q}{\Lambda}\right)^2
\left(\frac{M_{\rm UV}}{\Lambda}\right)^{25/3}
+\frac{3}{31}\left(\frac{y_Q^2}{16\pi^2}\right)^{10}
\left(\frac{M_{\rm UV}}{\Lambda}\right)^{31/3}\Bigg] \,.
\eeqa

For the sextet doublet $R_{11}^{(2)}$, the result is
\beqa
\label{eq:identical-sextet-doublet-uv}
\chi_{(6,2,1/6)}^{\rm UV}
&\sim&
2C_3^{(2)}(6,2)\left(\frac{2\pi}{\alpha_s(M_{\rm UV})}\right)^6
Y_{\rm SM}\Lambda_{\rm QCD}^{4}
\left(\frac{\Lambda_{\rm QCD}}{\Lambda}\right)^3\times\nl
& &\Bigg[
3\left(\frac{y_Q^2}{16\pi^2}\right)^5
\left(\frac{M_Q}{\Lambda}\right)^{10}
\left(\frac{M_{\rm UV}}{\Lambda}\right)^{1/3}
+\frac{3}{7}\left(\frac{y_Q^2}{16\pi^2}\right)^6
\left(\frac{M_Q}{\Lambda}\right)^8
\left(\frac{M_{\rm UV}}{\Lambda}\right)^{7/3}\nl
\ad\frac{3}{13}\left(\frac{y_Q^2}{16\pi^2}\right)^7
\left(\frac{M_Q}{\Lambda}\right)^6
\left(\frac{M_{\rm UV}}{\Lambda}\right)^{13/3}
+\frac{3}{19}\left(\frac{y_Q^2}{16\pi^2}\right)^8
\left(\frac{M_Q}{\Lambda}\right)^4
\left(\frac{M_{\rm UV}}{\Lambda}\right)^{19/3}\nl
\ad\frac{3}{25}\left(\frac{y_Q^2}{16\pi^2}\right)^9
\left(\frac{M_Q}{\Lambda}\right)^2
\left(\frac{M_{\rm UV}}{\Lambda}\right)^{25/3}
+\frac{3}{31}\left(\frac{y_Q^2}{16\pi^2}\right)^{10}
\left(\frac{M_{\rm UV}}{\Lambda}\right)^{31/3}\Bigg] \,.
\eeqa

Finally, the octet doublet $R_{13}^{(2)}$ yields the UV-sensitive susceptibility as
\beqa
\label{eq:identical-octet-doublet-uv}
\chi_{(8,2,-1/2)}^{\rm UV}
&\sim&
2C_3^{(2)}(8,2)\left(\frac{2\pi}{\alpha_s(M_{\rm UV})}\right)^6
Y_{\rm SM}\Lambda_{\rm QCD}^{4}
\left(\frac{\Lambda_{\rm QCD}}{\Lambda}\right)^3\times\nl
& &\Bigg[
\left(\frac{y_Q^2}{16\pi^2}\right)^6
\left(\frac{M_Q}{\Lambda}\right)^{12}
\left(\frac{M_{\rm UV}}{\Lambda}\right)
+\frac{1}{3}\left(\frac{y_Q^2}{16\pi^2}\right)^7
\left(\frac{M_Q}{\Lambda}\right)^{10}
\left(\frac{M_{\rm UV}}{\Lambda}\right)^3\nl
\ad\frac{1}{5}\left(\frac{y_Q^2}{16\pi^2}\right)^8
\left(\frac{M_Q}{\Lambda}\right)^8
\left(\frac{M_{\rm UV}}{\Lambda}\right)^5
+\frac{1}{7}\left(\frac{y_Q^2}{16\pi^2}\right)^9
\left(\frac{M_Q}{\Lambda}\right)^6
\left(\frac{M_{\rm UV}}{\Lambda}\right)^7\nl
\ad\frac{1}{9}\left(\frac{y_Q^2}{16\pi^2}\right)^{10}
\left(\frac{M_Q}{\Lambda}\right)^4
\left(\frac{M_{\rm UV}}{\Lambda}\right)^9
+\frac{1}{11}\left(\frac{y_Q^2}{16\pi^2}\right)^{11}
\left(\frac{M_Q}{\Lambda}\right)^2
\left(\frac{M_{\rm UV}}{\Lambda}\right)^{11}\nl
\ad\frac{1}{13}\left(\frac{y_Q^2}{16\pi^2}\right)^{12}
\left(\frac{M_{\rm UV}}{\Lambda}\right)^{13}\Bigg] \,.
\eeqa

In each above expressions, successive terms replace two mass insertions by one scalar-Yukawa loop, so that the final term carries the largest endpoint power. 
Among the perturbative identical-copy spectra, the strongest UV scaling is obtained for $R_{13}^{(2)}$, corresponding to two copies of the octet doublet $Q=(8,2,-1/2)$. 
The final term in Eq.~\eqref{eq:identical-octet-doublet-uv} scales as $\left(M_{\rm UV}/\Lambda\right)^{13}$. The next-largest power is $38/3$, obtained for $R_1^{(25)}$ and $R_9^{(5)}$.

\subsection{Distinct VQ sets}
\label{subsec:different-vqs}

We construct distinct VQ sets from the 15 VQs, given in Tab.~\ref{tab:vq-copy-number-bounds}, that satisfy the one-copy perturbativity requirement, allowing at most one copy of each representation. 
As in the previous subsection, all VQs within a given set are assigned a common mass $M_Q$ and a common Yukawa coupling $y_Q$. 
For each multiplicity $n$, we enumerate all $n$-member sets and discard any set for which at least one gauge coupling violates $\alpha_i(M_{\rm UV})<1$ at one loop. 
The surviving sets are then tested using the coupled two-loop gauge running. 
Among the allowed sets, we retain the one with the largest UV power $-\Delta$. 
If several sets have the same power, we select the one with the largest determinant prefactor $C_3$.

No set containing eight distinct representations satisfies the one-loop perturbativity condition for all three gauge factors. 
Therefore, at most seven distinct VQ representations can be included.\footnote{For a set $S$ of distinct VQs, Eq.~\eqref{eq:integrated-one-loop-alpha} is applied with
$n_Q\delta b_i\longrightarrow\sum_{R_a\in S}\delta b_i(R_a)$. The perturbativity condition in Eq.~\eqref{eq:continuous-copy-budget}, evaluated using the matching-scale couplings in Eq.~\eqref{eq:matched-gauge-couplings}, then gives
\[
15\sum_{R_a\in S}\delta b_1(R_a)\leq179,\quad
3\sum_{R_a\in S}\delta b_2(R_a)\leq53,\quad
3\sum_{R_a\in S}\delta b_3(R_a)\leq57.
\]
The factors $15$ and $3$ convert the rational entries of $\delta b$ in Tab.~\ref{tab:vq-copy-number-bounds} into integers. Among the eight-member subsets of the 15 VQs satisfying the one-copy perturbativity requirement, the smallest value of $3\sum\delta b_3(R_a)$ compatible with the first two inequalities is $58$. 
It exceeds the ${\rm SU}(3)_c$ limit of $57$. 
Therefore, no set of eight distinct VQs satisfies all three conditions above.} 
The sets selected by this procedure are listed in Tab.~\ref{tab:distinct-maximal-uv-sets}.
\begin{table*}[t!]
\centering
\small
\renewcommand{\arraystretch}{1.00}
\setlength{\tabcolsep}{3.5pt}
\begin{tabular}{c l c c c c c c}
\hline\hline
$n$ & $S_n$ & $2N_S$ & $(k,r)$ & $-\Delta$ & $C_3(S_n)$
& $\big(E/N,N_{\rm DW}\big)_{\rm min}$
& $\big(E/N,N_{\rm DW}\big)_{\rm max}$\\
\hline
$2$ & $\{R_{11},R_{13}\}$ & $22$ & $(11,0)$ & $35/3$ & $1.75$ & $(14/3,2)$ & $(34/33,22)$\\
$3$ & $\{R_3,R_{11},R_{13}\}$ & $24$ & $(12,0)$ & $13$ & $3.13$ & $(19/6,4)$ & $(13/12,24)$\\
$4$ & \begin{tabular}[c]{@{}l@{}}$\{R_1,R_2$\\$R_{11},R_{13}\}$\end{tabular}
& $24$ & $(12,0)$ & $13$ & $3.13$ & $(11/3,2)$ & $(31/33,22)$\\
$5$ & \begin{tabular}[c]{@{}l@{}}$\{R_1,R_3,R_6$\\$R_{12},R_{13}\}$\end{tabular}
& $24$ & $(12,0)$ & $13$ & $4.14$ & $(5/3,6)$ & $(-5/3,6)$\\
$6$ & \begin{tabular}[c]{@{}l@{}}$\{R_1,R_2,R_3$\\$R_6,R_9,R_{13}\}$\end{tabular}
& $24$ & $(12,0)$ & $13$ & $4.14$ & $(-1/3,2)$ & $(7/24,16)$\\
$7$ & \begin{tabular}[c]{@{}l@{}}$\{R_1,R_2,R_3,R_4$\\$R_6,R_9,R_{11}\}$\end{tabular}
& $24$ & $(12,0)$ & $13$ & $4.14$ & $(5/3,4)$ & $(-4/3,12)$\\
\hline\hline
\end{tabular}
\caption{Sets of distinct VQs that maximize the UV power at fixed
multiplicity while remaining gauge perturbative at two loops.  The matching
conditions are $\Lambda=5\times10^{11}\,\GeV$ and $y_Q=1$.  For each set, $2N_S=\sum_a2N_a$.  The last two columns are obtained by computing $E_S=\sum_a\Delta X_aE_a$, $N_S=\sum_a\Delta X_aN_a$ and $N_{\rm DW}=2|N_S|$, and scanning $\Delta X_a=\pm1$ with $N_S\neq0$. They list the assignments that minimize and maximize $N_{\rm DW}|E/N-1.92|$, respectively, and hence extremize the axion--photon coupling at fixed $\Lambda$. In each ordered pair, the displayed $N_{\rm DW}$ belongs to the corresponding PQ-charge assignment.  The individual values of $E_a$ and $N_a$ are given in Tab.~\ref{tab:vq-copy-number-bounds}.}
\label{tab:distinct-maximal-uv-sets}
\end{table*}
The corresponding leading UV sensitive instanton contributions of their susceptibility are
\beqa
\label{eq:distinct-maximal-uv-contributions}
\chi_{S_2}^{\rm UV}
&\sim&
2C_3(S_2)\left(\frac{2\pi}{\alpha_s(M_{\rm UV})}\right)^6
Y_{\rm SM}\Lambda_{\rm QCD}^{4}
\left(\frac{\Lambda_{\rm QCD}}{\Lambda}\right)^3
\frac{3}{35}\left(\frac{y_Q^2}{16\pi^2}\right)^{11}
\left(\frac{M_{\rm UV}}{\Lambda}\right)^{35/3}\,,\nl
\chi_{S_3}^{\rm UV}
&\sim&
2C_3(S_3)\left(\frac{2\pi}{\alpha_s(M_{\rm UV})}\right)^6
Y_{\rm SM}\Lambda_{\rm QCD}^{4}
\left(\frac{\Lambda_{\rm QCD}}{\Lambda}\right)^3
\frac{1}{13}\left(\frac{y_Q^2}{16\pi^2}\right)^{12}
\left(\frac{M_{\rm UV}}{\Lambda}\right)^{13}\,,\nl
\chi_{S_4}^{\rm UV}
&\sim&
2C_3(S_4)\left(\frac{2\pi}{\alpha_s(M_{\rm UV})}\right)^6
Y_{\rm SM}\Lambda_{\rm QCD}^{4}
\left(\frac{\Lambda_{\rm QCD}}{\Lambda}\right)^3
\frac{1}{13}\left(\frac{y_Q^2}{16\pi^2}\right)^{12}
\left(\frac{M_{\rm UV}}{\Lambda}\right)^{13}\,,\nl
\chi_{S_5}^{\rm UV}
&\sim&
2C_3(S_5)\left(\frac{2\pi}{\alpha_s(M_{\rm UV})}\right)^6
Y_{\rm SM}\Lambda_{\rm QCD}^{4}
\left(\frac{\Lambda_{\rm QCD}}{\Lambda}\right)^3
\frac{1}{13}\left(\frac{y_Q^2}{16\pi^2}\right)^{12}
\left(\frac{M_{\rm UV}}{\Lambda}\right)^{13}\,,\nl
\chi_{S_6}^{\rm UV}
&\sim&
2C_3(S_6)\left(\frac{2\pi}{\alpha_s(M_{\rm UV})}\right)^6
Y_{\rm SM}\Lambda_{\rm QCD}^{4}
\left(\frac{\Lambda_{\rm QCD}}{\Lambda}\right)^3
\frac{1}{13}\left(\frac{y_Q^2}{16\pi^2}\right)^{12}
\left(\frac{M_{\rm UV}}{\Lambda}\right)^{13}\,,\nl
\chi_{S_7}^{\rm UV}
&\sim&
2C_3(S_7)\left(\frac{2\pi}{\alpha_s(M_{\rm UV})}\right)^6
Y_{\rm SM}\Lambda_{\rm QCD}^{4}
\left(\frac{\Lambda_{\rm QCD}}{\Lambda}\right)^3
\frac{1}{13}\left(\frac{y_Q^2}{16\pi^2}\right)^{12}
\left(\frac{M_{\rm UV}}{\Lambda}\right)^{13}\,.
\eeqa

For a single VQ, UV-sensitive contributions occur for the  $R_{11,12,13,14,16,17}$. The identical-copy spectra with the dominant UV-sensitive contributions are $R_1^{(25)}$, $R_2^{(11)}$, $R_3^{(8)}$, $R_4^{(3)}$, $R_6^{(2)}$, $R_7^{(2)}$, $R_9^{(5)}$, $R_{10}^{(4)}$, $R_{11}^{(2)}$, and $R_{13}^{(2)}$. The largest power encountered in these susceptibilities is $(M_{\rm UV}/\Lambda)^{13}$, attained by the identical $R_{13}^{(2)}$ and by the distinct VQ sets $S_3,\ldots,S_7$. Among the remaining identical-copy cases, the largest exponent is $38/3$, obtained for $R_1^{(25)}$ and $R_9^{(5)}$, whereas the largest single-VQ exponent is $31/3$, obtained for $R_{16,17}$. 

\section{Implications for axion phenomenology}
\label{sec:axion-pheno}

In Sec.~\ref{sec:vq-models} we determined the UV sensitivity of the instanton contribution for one VQ, multiple identical VQs and sets of different VQs. 
We now examine how these contributions modify the axion mass and its coupling to photons.

In general, the QCD and UV contributions can carry different phases and need not select the same vacuum. 
We consider the case in which these two potentials have the same periodicity and the same CP-conserving minimum. 
Their susceptibilities then add without shifting the minimum.
By applying Eq.~\eqref{eq:instanton-axion-potential} with $\chi=\chiQCD+\chi_{\rm SI}$, the combined axion potential is
\be
\label{eq:aligned-axion-potential}
V(a)=-\left(\chiQCD+\chi_{\rm SI}\right)
\cos\left(\bar\theta+\frac{a}{f_a}\right)\,,
\ee
where $\chi_{\rm SI}$ denotes the small-instanton susceptibility calculated for various cases in Sec.~\ref{sec:vq-models}. 
By using Eq.~\eqref{eq:susceptibility-and-axion-mass}, the mass generated by QCD alone and the physical axion mass due to the small instanton are
\be
\label{eq:qcd-si-axion-masses}
m_{a,{\rm QCD}}^2=\frac{\chiQCD}{f_a^2},
\quad
m_a^2=\frac{\chiQCD+\chi_{\rm SI}}{f_a^2}\,,
\ee
respectively.
Consequently, the relative change in the mass squared is
\be
\label{eq:susceptibility-to-mass-squared}
\frac{\chi_{\rm SI}}{\chiQCD}
=\frac{m_a^2-m_{a,{\rm QCD}}^2}{m_{a,{\rm QCD}}^2}\,.
\ee
For the numerical analysis, we use that
\be
\label{eq:qcd-susceptibility-value}
\chiQCD^{1/4}=75.44(34)\,{\rm MeV}\,,
\quad
\chiQCD=3.24\times10^{-5}\,\GeV^4\,,
\ee
with the known higher-order corrections~\cite{GrillidiCortona:2015jxo,Gorghetto:2018ocs} included. 
For the SM zero-mode factor, we use the value of $Y_{\rm SM}$ defined in Eq.~\eqref{eq:sm-zero-mode-factor} and the associated RG effect is not taken into account. 
In contrast, the VQ Yukawa coupling of $y_Q$ is specified at the matching scale $\Lambda$, and its running above $\Lambda$ is also not included. 

Although the matching scale $\Lambda=m_\sigma$ is varied in the plots, the maximum identical-copy multiplicities of different VQs and the distinct-VQ sets are those selected at the reference value $\Lambda=5\times10^{11}\,\GeV$. 
We therefore compare the same spectra as $\Lambda$ is changed, rather than recalculating the allowed VQ content at every matching scale. 
We do not extend the scan below $\Lambda=5\times10^{10}\,\GeV$.\footnote{For $\lambda_\Phi=1$, $f_a=\Lambda/(\sqrt{2}N_{\rm DW})$. 
The largest domain-wall number appearing in our cases is $N_{\rm DW}=25$, for which $\Lambda=5\times10^{10}\,\GeV$ gives $f_a=1.41\times10^9\,\GeV$. 
Consequently, further lowering $\Lambda$ can therefore place some spectra below the conservative astrophysical benchmark $f_a\gtrsim10^9\,\GeV$~\cite{DiLuzio:2020wdo,ChangSN1987A}.} 
The mass plots use three values of $\Lambda$ stated in their panels and scan $y_Q$ through $M_Q$, whereas the axion--photon plots use the fixed $y_Q=1$ and the varying $\Lambda$'s. 

\subsection{Axion mass}

\begin{figure*}[t!]
\centering
\begin{minipage}{0.328\textwidth}
\centering
\includegraphics[width=\linewidth]{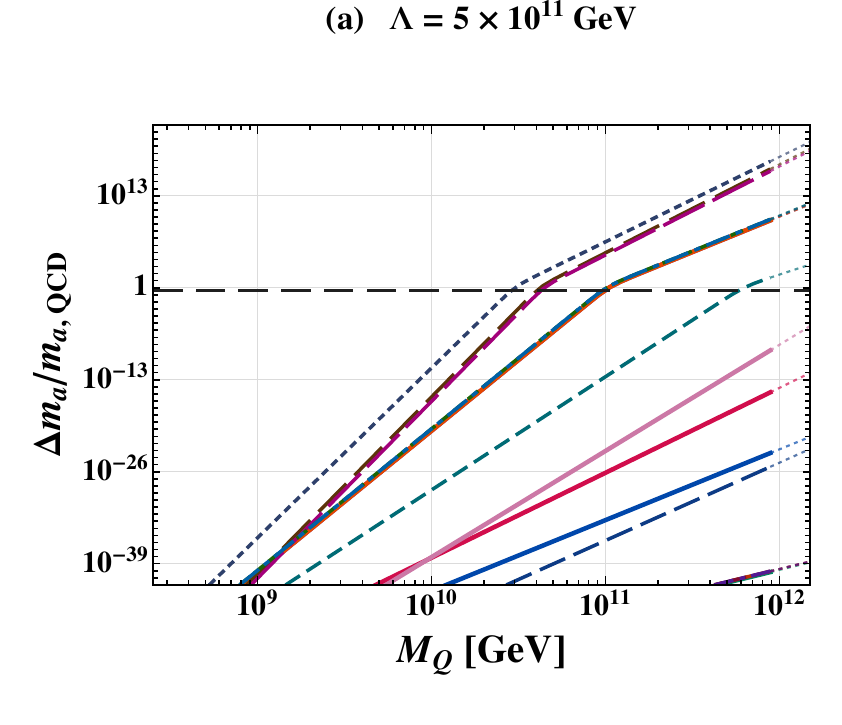}
\end{minipage}\hfill
\begin{minipage}{0.328\textwidth}
\centering
\includegraphics[width=\linewidth]{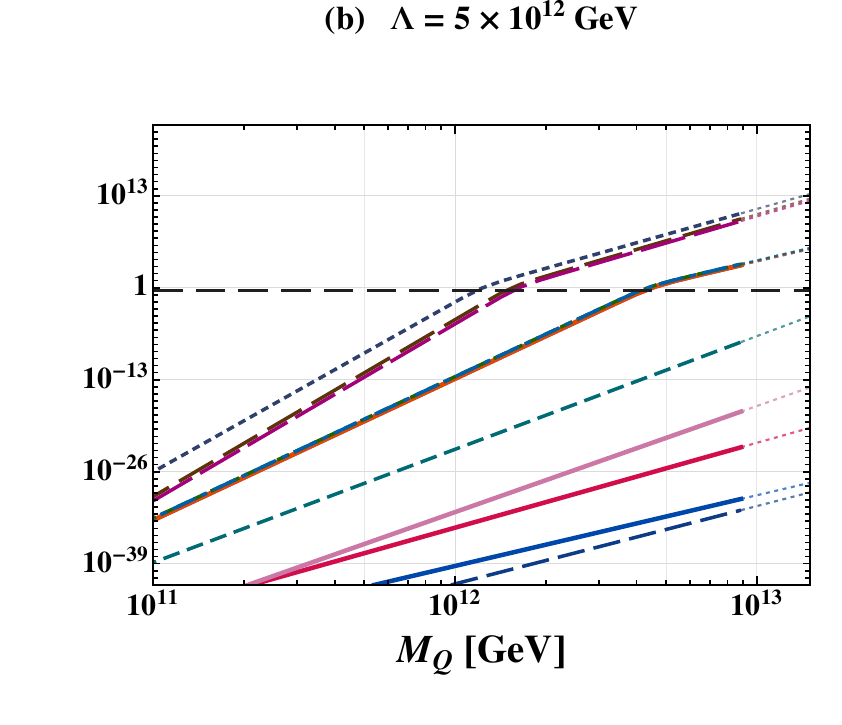}
\end{minipage}\hfill
\begin{minipage}{0.328\textwidth}
\centering
\includegraphics[width=\linewidth]{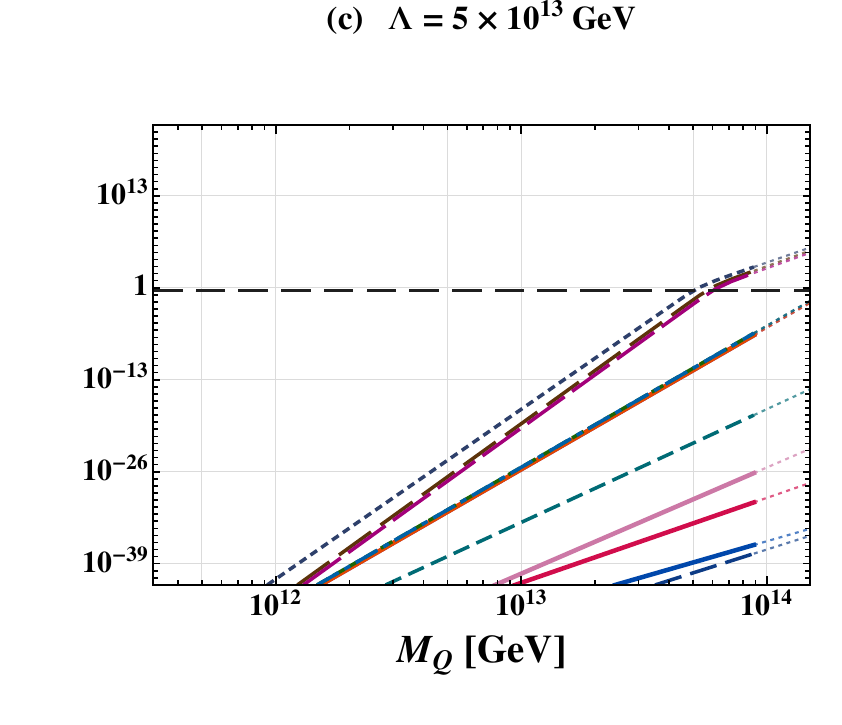}
\end{minipage}
\caption{Axion-mass shift as a function of the VQ mass at
(a) $\Lambda=5\times10^{11}\,\GeV$,
(b) $\Lambda=5\times10^{12}\,\GeV$, and
(c) $\Lambda=5\times10^{13}\,\GeV$.
Solid curves denote one VQ, while the patterned dashed curves denote the maximal spectra made from identical copies, and the thin dotted extensions points when $y_Q>\sqrt{4\pi}$.
The one-VQ curves are $R_{11}$~(\figSolidKey{0,0.28,0.67}),
$R_{12}$~(\figSolidKey{0,0.52,0.38}),
$R_{13}$~(\figSolidKey{0.82,0.05,0.30}),
$R_{14}$~(\figSolidKey{0.80,0.47,0.65}),
$R_{16}$~(\figSolidKey{0.46,0.16,0.62}), and
$R_{17}$~(\figSolidKey{0.90,0.28,0}).
The maximal identical-copy spectra are
$R_1^{(25)}$~(\figLongDashKey{0.36,0.20,0.04}),
$R_2^{(11)}$~(\figLongDashKey{0.05,0.23,0.52}),
$R_3^{(8)}$~(\figLongDashKey{0,0.42,0.46}),
$R_4^{(3)}$~(\figLongDashKey{0.73,0.19,0.02}),
$R_6^{(2)}$~(\figLongDashKey{0.50,0.04,0.19}),
$R_7^{(2)}$~(\figLongDashKey{0.32,0.10,0.58}),
$R_9^{(5)}$~(\figLongDashKey{0.64,0,0.48}),
$R_{10}^{(4)}$~(\figLongDashKey{0.10,0.40,0.03}),
$R_{11}^{(2)}$~(\figLongDashKey{0,0.38,0.66}), and
$R_{13}^{(2)}$~(\figLongDashKey{0.18,0.25,0.42}).
The black dashed horizontal line marks~(\figLongDashKey{0,0,0}) the equal susceptibilities of $\chi_{\rm SI}=\chiQCD$.}
\label{fig:mass-shift-single-identical}
\end{figure*}

The additional susceptibility increases the curvature of the potential and therefore raises the physical axion mass. 
By defining $\Delta m_a\equiv m_a-m_{a,{\rm QCD}}$, the exact fractional shift in the axion mass is
\be
\label{eq:exact-aligned-mass-shift}
\frac{\Delta m_a}{m_{a,{\rm QCD}}}
=\sqrt{1+\frac{\chi_{\rm SI}}{\chiQCD}}-1
=\frac{\chi_{\rm SI}/\chiQCD}
{\sqrt{1+\chi_{\rm SI}/\chiQCD}+1}\,.
\ee
The limiting forms of fractional shift in the axion mass, given in Eq.~\eqref{eq:exact-aligned-mass-shift}, are
\be
\label{eq:aligned-mass-shift-limits}
\frac{\Delta m_a}{m_{a,{\rm QCD}}}\simeq
\begin{cases}
\dfrac{1}{2}\dfrac{\chi_{\rm SI}}{\chiQCD}, & \hspace{1cm} \dfrac{\chi_{\rm SI}}{\chiQCD}\ll1 \,,\\[6pt]
\sqrt{\dfrac{\chi_{\rm SI}}{\chiQCD}}, & \hspace{1cm}  \dfrac{\chi_{\rm SI}}{\chiQCD}\gg1 \,.
\end{cases}
\ee
 Equal susceptibilities of $\chi_{\rm SI}=\chiQCD$ give $\Delta m_a/m_{a,{\rm QCD}}=\sqrt{2}-1\simeq0.41$.
 Above this value the small-instanton term is the leading source of the axion mass.
\begin{figure*}[t!]
\centering
\begin{minipage}{0.328\textwidth}
\centering
\includegraphics[width=\linewidth]{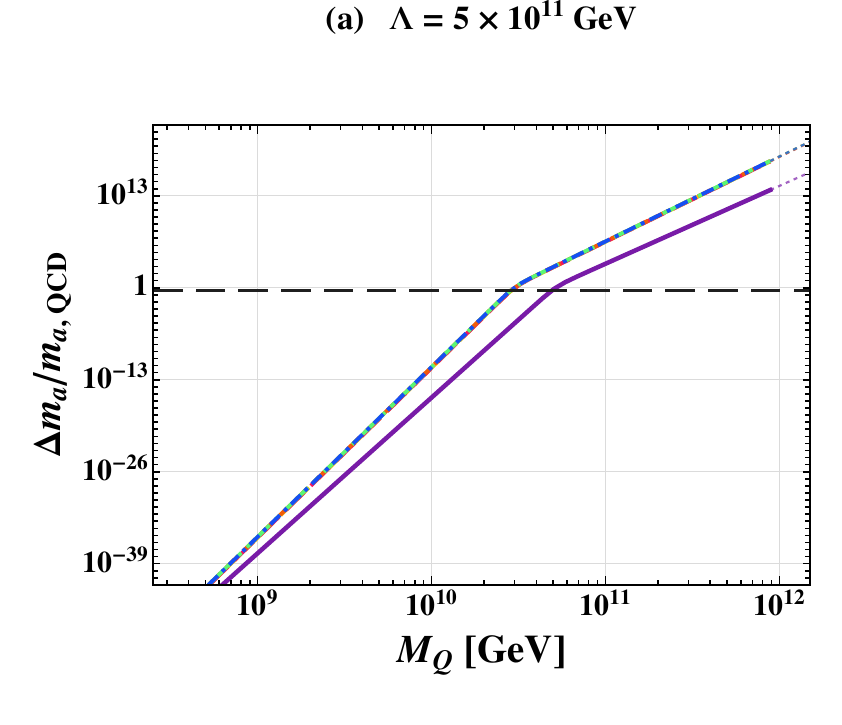}
\end{minipage}\hfill
\begin{minipage}{0.328\textwidth}
\centering
\includegraphics[width=\linewidth]{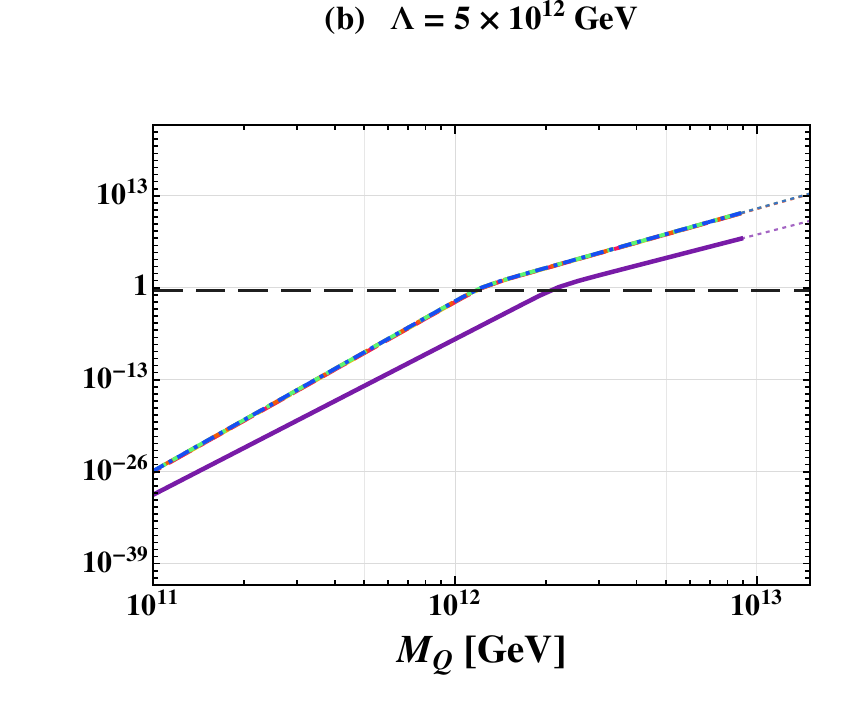}
\end{minipage}\hfill
\begin{minipage}{0.328\textwidth}
\centering
\includegraphics[width=\linewidth]{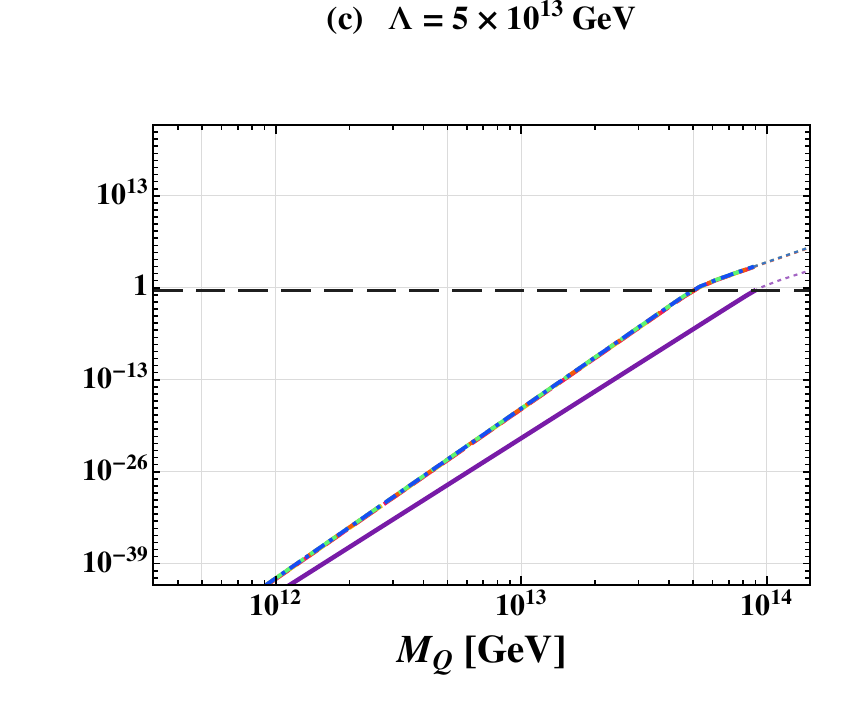}
\end{minipage}
\caption{Axion-mass shift for the distinct-VQ spectra, for the same values of $\Lambda$ as in Fig.~\ref{fig:mass-shift-single-identical}. The $S_2$ contour~(\figSolidKey{0.471,0.106,0.654}) is shown separately; the $S_3,\ldots,S_7$ contours overlap on this scale and are represented collectively by the visible $S_7$ contour~(\figLongDashKey{0.088,0.304,0.952}). All other conventions are as in Fig.~\ref{fig:mass-shift-single-identical}.}
\label{fig:mass-shift-distinct-multiplicity}
\end{figure*}

Fig.~\ref{fig:mass-shift-single-identical} compares the UV-dominated contributions represented by the solid one-VQ contours with those represented by the patterned dashed contours for the maximal identical-copy spectra as a function of mass insertions $M_Q$. 
Panels (a), (b), and (c) correspond to fixed matching scales $\Lambda=5\times10^{11}\,\GeV$, $5\times10^{12}\,\GeV$, and $5\times10^{13}\,\GeV$, respectively. 
The one-VQ contours include $R_{14}$~(\figSolidKey{0.80,0.47,0.65}), $R_{16}$~(\figSolidKey{0.46,0.16,0.62}), and $R_{17}$~(\figSolidKey{0.90,0.28,0}), while $R_{13}^{(2)}$~(\figLongDashKey{0.18,0.25,0.42}) is an identical-copy contour while the remaining color assignments are given in the caption of Fig.~\ref{fig:mass-shift-single-identical}. 
For the one-VQ spectra, the plotted UV contributions are taken from Eqs.~\eqref{eq:vq15-ir-uv-contributions}, \eqref{eq:vq-decuplet-singlet-ir-log-uv}, and~\eqref{eq:vq-sextet-doublet-ir-uv}--\eqref{eq:vq-octet-doublet-ir-uv}, while the identical-copy contributions are taken from Eqs.~\eqref{eq:identical-triplet-singlet-minus-uv}--\eqref{eq:identical-octet-doublet-uv}. 
For a closure specified by $(k,r)$ with $\Delta<0$, Eq.~\eqref{eq:instanton-susceptibility-endpoints} gives the following parametric dependence of 
\be
\label{eq:uv-susceptibility-plot-scaling}
\chi_{\rm SI}
\approx \Lambda^4_{\rm QCD}\,
\left(\frac{2\pi}{\alpha_s(M_{\rm UV})}\right)^6
\left(\frac{\Lambda_{\rm QCD}}{\Lambda}\right)^3
\left(\frac{M_Q}{\Lambda}\right)^r
\left(\frac{y_Q^2}{16\pi^2}\right)^k
\frac{1}{-\Delta}
\left(\frac{M_{\rm Pl}}{\Lambda}\right)^{-\Delta}\,.
\ee
The factors $(M_Q/\Lambda)^r$ and $(y_Q^2/16\pi^2)^k$ account for the remaining mass insertions and scalar--Yukawa loops, respectively. 
The factor $(M_{\rm UV}/\Lambda)^{-\Delta}$ describes the enhancement generated by the UV endpoint of the instanton-size integral, where the gauge prefactor is evaluated at $\alpha_s(M_{\rm UV})$. 
With $\Lambda=m_\sigma$ and $\lambda_\Phi=1$, Eq.~\eqref{eq:radial-and-vq-masses} gives $y_Q=2M_Q/\Lambda$. 
Moving along a contour by varying $M_Q$ therefore changes both the mass-insertion and Yukawa-loop factors. 
When $r=0$, the $M_Q$ dependence enters entirely through $y_Q$. 
The resulting susceptibility is normalized using Eq.~\eqref{eq:qcd-susceptibility-value} and converted into the exact fractional mass shift through Eq.~\eqref{eq:exact-aligned-mass-shift} and plotted for different cases. 
The thin dotted extensions lie beginning at $M_Q=\sqrt{\pi}\Lambda$, or equivalently $y_Q=\sqrt{4\pi}$ denotes the contours for non-perturbative Yukawas. 
Increasing $\Lambda$ suppresses both the common matching factor $(\Lambda_{\rm QCD}/\Lambda)^3$ and the UV enhancement $(M_{\rm UV}/\Lambda)^{-\Delta}$, while shifting the Yukawa-perturbativity boundary toward larger $M_Q$.

Fig.~\ref{fig:mass-shift-distinct-multiplicity} shows the same construction but for the distinct-VQ susceptibilities provided in Eq.~\eqref{eq:distinct-maximal-uv-contributions}. The $S_2$ contour~(\figSolidKey{0.471,0.106,0.654}) has $(k,r)=(11,0)$ and $-\Delta=35/3$, whereas $S_3,\ldots,S_7$ have $(k,r)=(12,0)$ and $-\Delta=13$. The latter sets consequently have the same dependence on $M_Q$ and on the UV ratio $M_{\rm UV}/\Lambda$. Their determinant prefactors differ only by order-one factors, so their curves overlap on the logarithmic scale of the figure. The bends in Figs.~\ref{fig:mass-shift-single-identical} and \ref{fig:mass-shift-distinct-multiplicity}, near $\chi_{\rm SI}=\chiQCD$, arise from the relation in Eq.~\eqref{eq:exact-aligned-mass-shift}, while the limiting behavior on either side is given by Eq.~\eqref{eq:aligned-mass-shift-limits}.

 For a single VQ, $R_{16}$ and $R_{17}$ give the largest axion mass of $m_a=3.43\,{\rm eV}$. 
 For identical copies, $R_{13}^{(2)}$ reaches $m_a =71.2\,{\rm MeV}$. 
 For distinct VQ sets, the largest physical mass is $m_a=65.3\,{\rm MeV}$ for the $N_{\rm DW}=22$ assignment of $S_4$. 
 Although $S_5$, $S_6$ and $S_7$ have the largest fractional shifts, their smaller domain-wall numbers give smaller physical masses. 
 Thus, additional identical or distinct VQs raise the largest mass from the eV scale to tens of MeV. 
 All above axion masses are quoted at $\Lambda = 5\times 10^{11}$ GeV and $y_Q=1$.

\subsection{Axion--photon coupling}
\label{subsec:axion-photon-coupling}

At energies below the QCD scale the axion--photon coupling is computed through chiral perturbation theory and is given as~\cite{GrillidiCortona:2015jxo,DiLuzio:2017pfr,Brandt:2026wir}
\be
\label{eq:axion-photon-coupling}
g_{a\gamma\gamma}
=\frac{\alpha_{\rm em}}{2\pi f_a}
\left(\frac{E}{N}-1.92\right)\,,
\ee
where $E/N$ is the electromagnetic-to-color anomaly ratio~(cf. Eq.~\eqref{eq:vq-anomaly-coefficients}) and $\alpha_{\rm em}$ is the low-energy electromagnetic coupling. 
The aligned small-instanton contribution changes the mass but not $g_{a\gamma\gamma}$ at fixed $f_a$ and anomaly coefficients. 
Eliminating $f_a$ in Eq.~\eqref{eq:axion-photon-coupling} through Eq.~\eqref{eq:qcd-si-axion-masses}, where $m_a$ is the physical mass generated by the combined susceptibility $\chiQCD+\chi_{\rm SI}$, gives
\be
\label{eq:axion-photon-mass-with-small-instantons}
g_{a\gamma\gamma}
=\frac{\alpha_{\rm em}}{2\pi}
\left(\frac{E}{N}-1.92\right)
\frac{m_a}{\sqrt{\chiQCD}}
\left(1+\frac{\chi_{\rm SI}}{\chiQCD}\right)^{-1/2}\,.
\ee
\begin{figure*}[t!]
\centering
\includegraphics[width=0.88\textwidth]{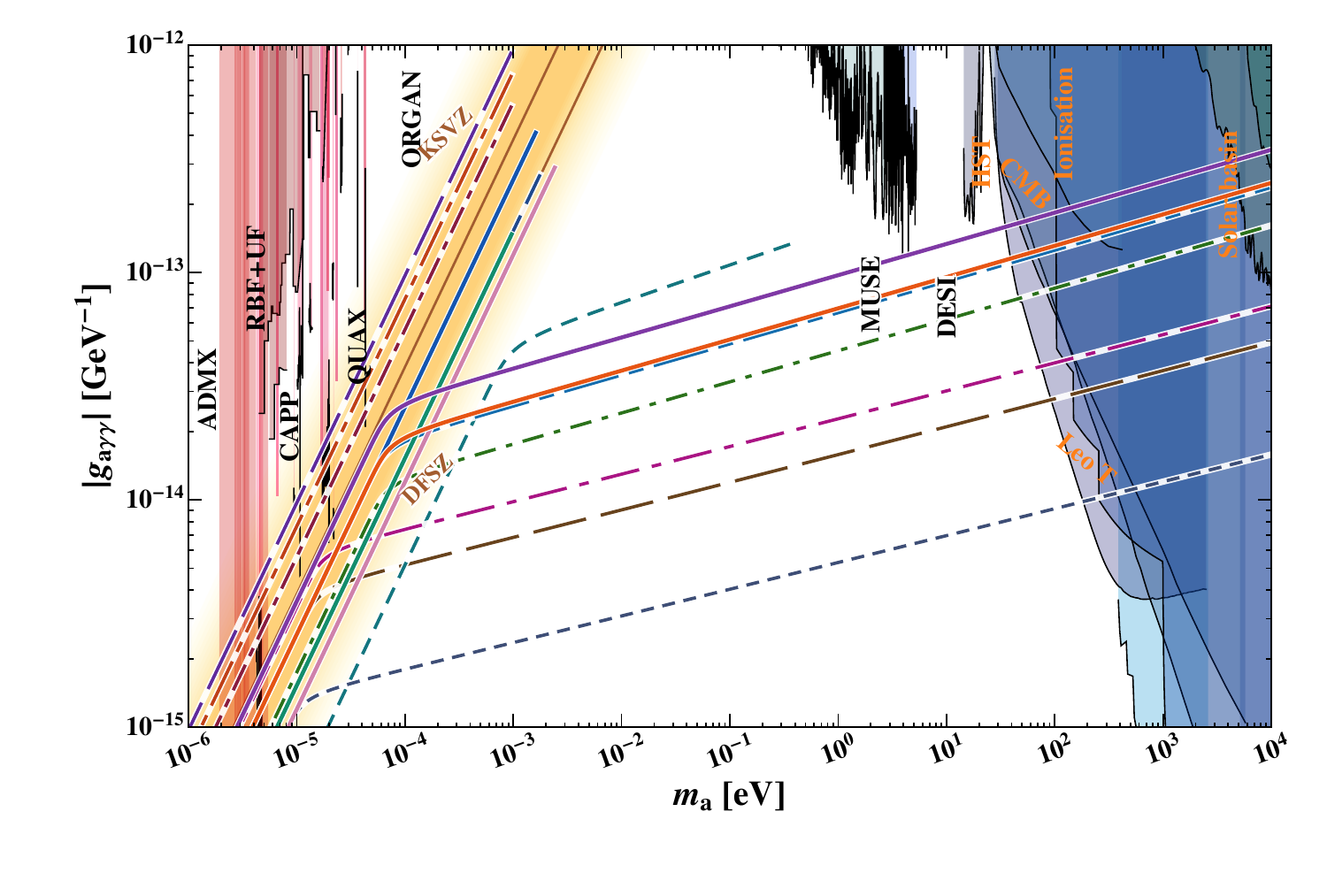}
\caption{The plot shows the variation of absolute axion--photon coupling as a function of the physical axion mass for one VQ and for the maximal identical-VQ spectra, with $y_Q=1$ and $5\times10^{10}\leq\Lambda/\GeV\leq5\times10^{16}$. Solid curves denote one VQ: $R_{11}$~(\figSolidKey{0,0.28,0.67}), $R_{12}$~(\figSolidKey{0,0.52,0.38}), $R_{13}$~(\figSolidKey{0.82,0.05,0.30}), $R_{14}$~(\figSolidKey{0.80,0.47,0.65}), $R_{16}$~(\figSolidKey{0.46,0.16,0.62}), and $R_{17}$~(\figSolidKey{0.90,0.28,0}). Patterned dashed curves denote the maximal identical-copy spectra: $R_1^{(25)}$~(\figLongDashKey{0.36,0.20,0.04}), $R_2^{(11)}$~(\figLongDashKey{0.05,0.23,0.52}), $R_3^{(8)}$~(\figLongDashKey{0,0.42,0.46}), $R_4^{(3)}$~(\figLongDashKey{0.73,0.19,0.02}), $R_6^{(2)}$~(\figLongDashKey{0.50,0.04,0.19}), $R_7^{(2)}$~(\figLongDashKey{0.32,0.10,0.58}), $R_9^{(5)}$~(\figLongDashKey{0.64,0,0.48}), $R_{10}^{(4)}$~(\figLongDashKey{0.10,0.40,0.03}), $R_{11}^{(2)}$~(\figLongDashKey{0,0.38,0.66}), and $R_{13}^{(2)}$~(\figLongDashKey{0.18,0.25,0.42}). The yellow region~(\figPatchKey{0.64,0.36,0.18}{0.98,0.91,0.64}) is the conventional QCD-axion band. Experimental, astrophysical and cosmological limits are taken from AxionLimits~\cite{AxionLimits}.}
\label{fig:axion-photon-lambda-scan}
\end{figure*}
Using the Thomson-limit value $\alpha_{\rm em}(0)^{-1}=137.04$ and Eq.~\eqref{eq:qcd-susceptibility-value}, the expression containing  $g_{a\gamma\gamma}$ becomes
\be
\label{eq:axion-photon-mass-numerical}
g_{a\gamma\gamma}
=2.04\times10^{-10}\,\GeV^{-1}
\left(\frac{m_a}{{\rm eV}}\right)
\left(\frac{E}{N}-1.92\right)
\left(1+\frac{\chi_{\rm SI}}{\chiQCD}\right)^{-1/2} \,.
\ee

\begin{figure*}[t!]
\centering
\begin{minipage}{0.49\textwidth}
\centering
\includegraphics[width=\linewidth]{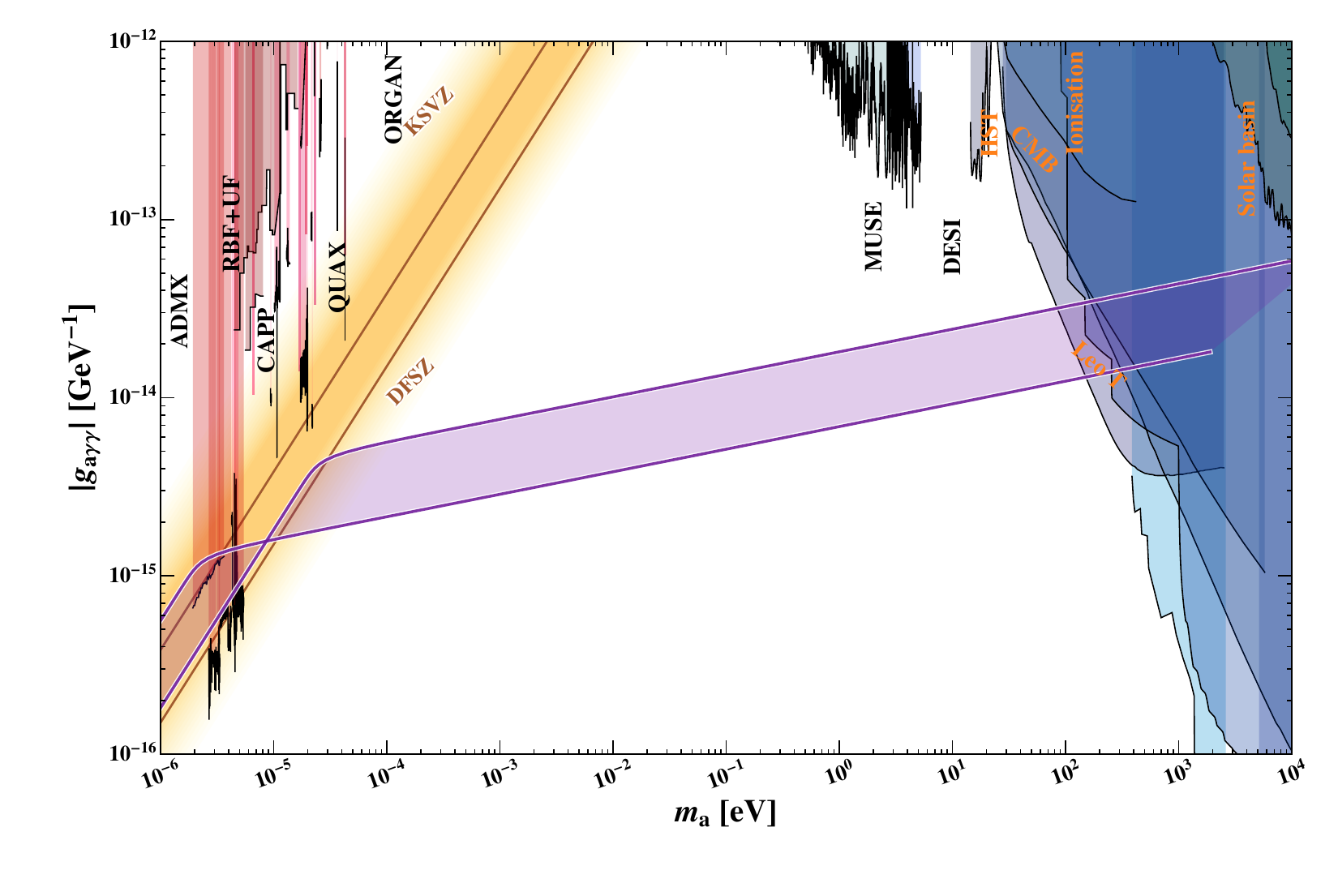}
\par\vspace{-0.6em}\small (a) $S_2$
\end{minipage}\hfill
\begin{minipage}{0.49\textwidth}
\centering
\includegraphics[width=\linewidth]{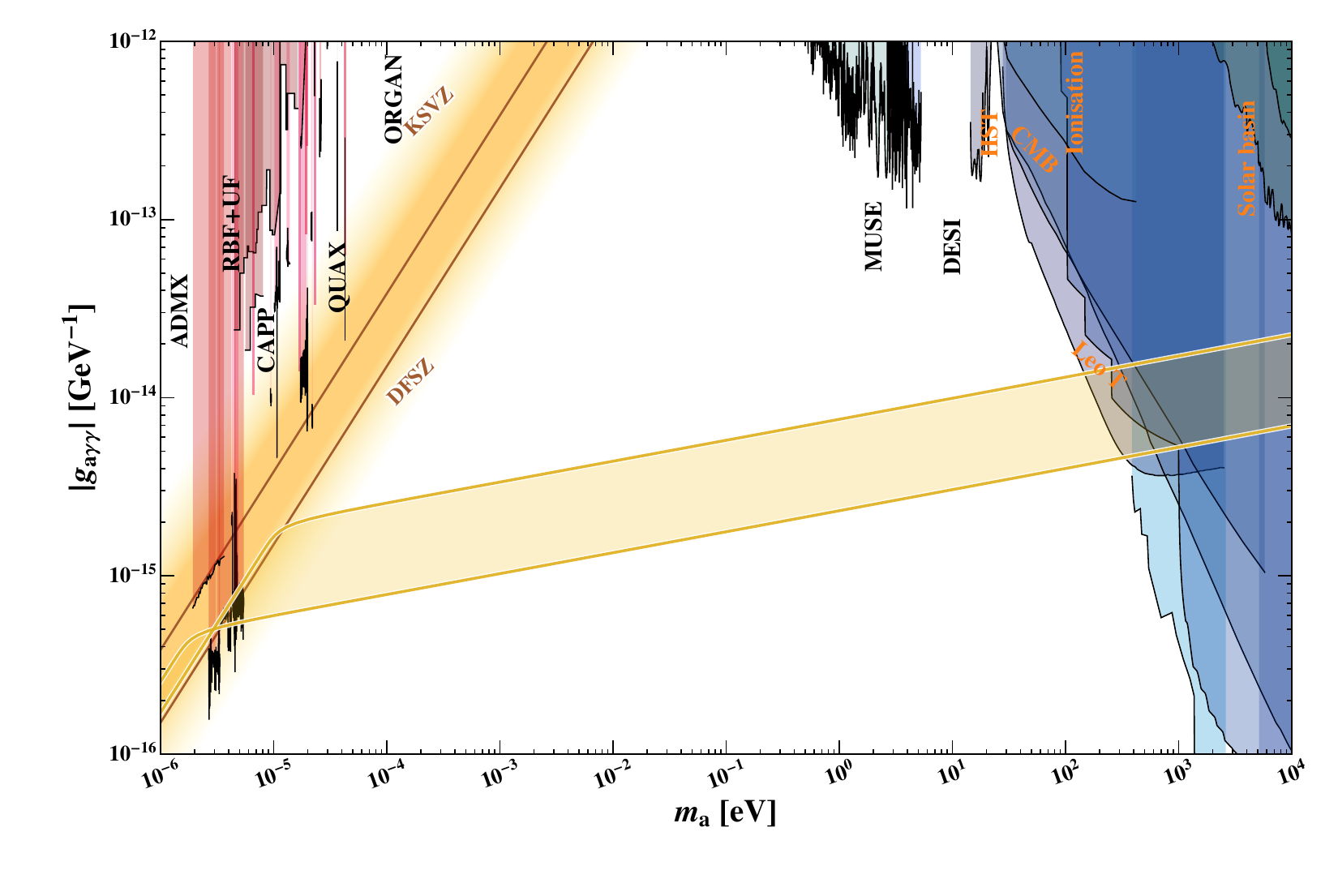}
\par\vspace{-0.6em}\small (b) $S_3$
\end{minipage}
\vspace{0.35em}

\begin{minipage}{0.49\textwidth}
\centering
\includegraphics[width=\linewidth]{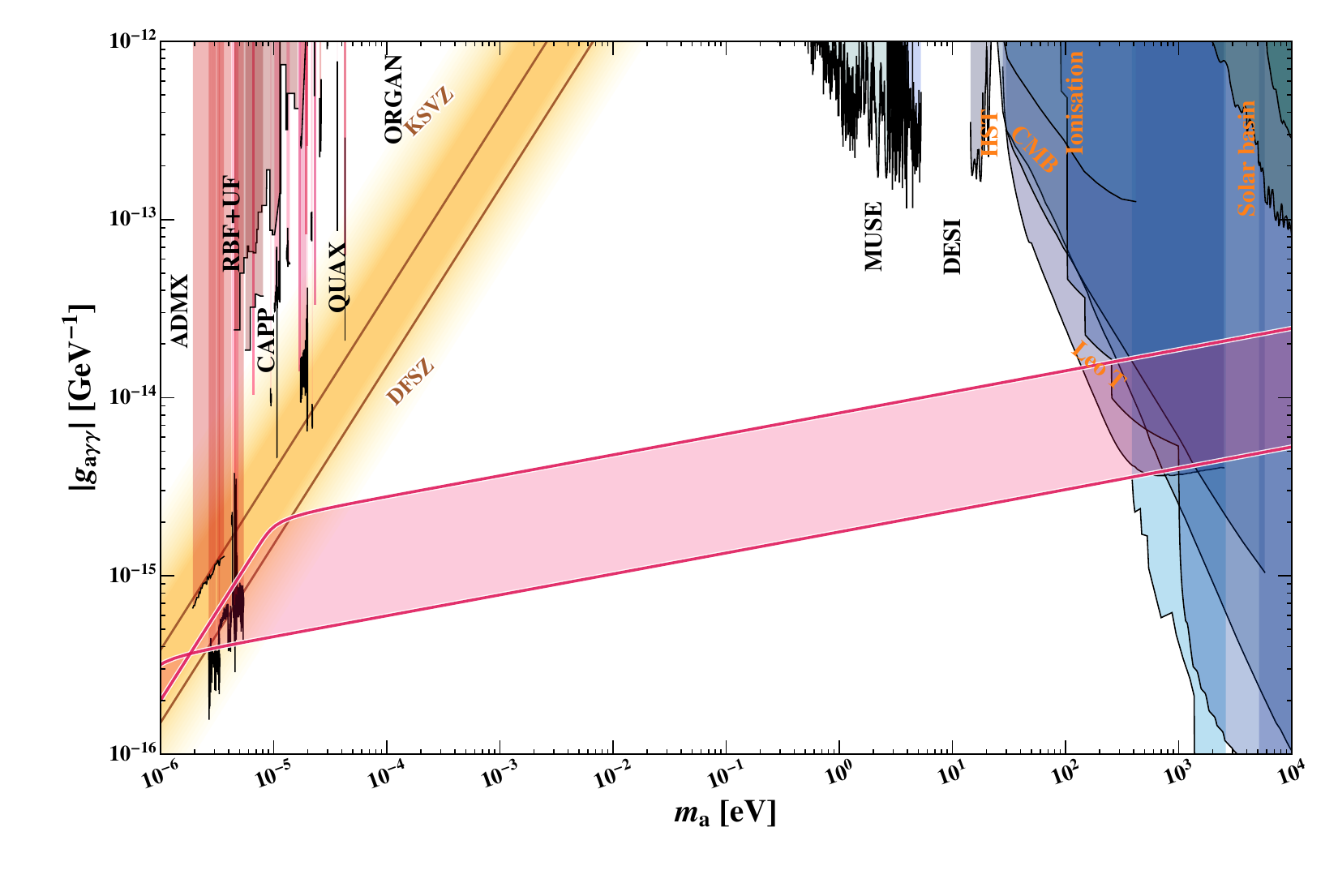}
\par\vspace{-0.6em}\small (c) $S_4$
\end{minipage}\hfill
\begin{minipage}{0.49\textwidth}
\centering
\includegraphics[width=\linewidth]{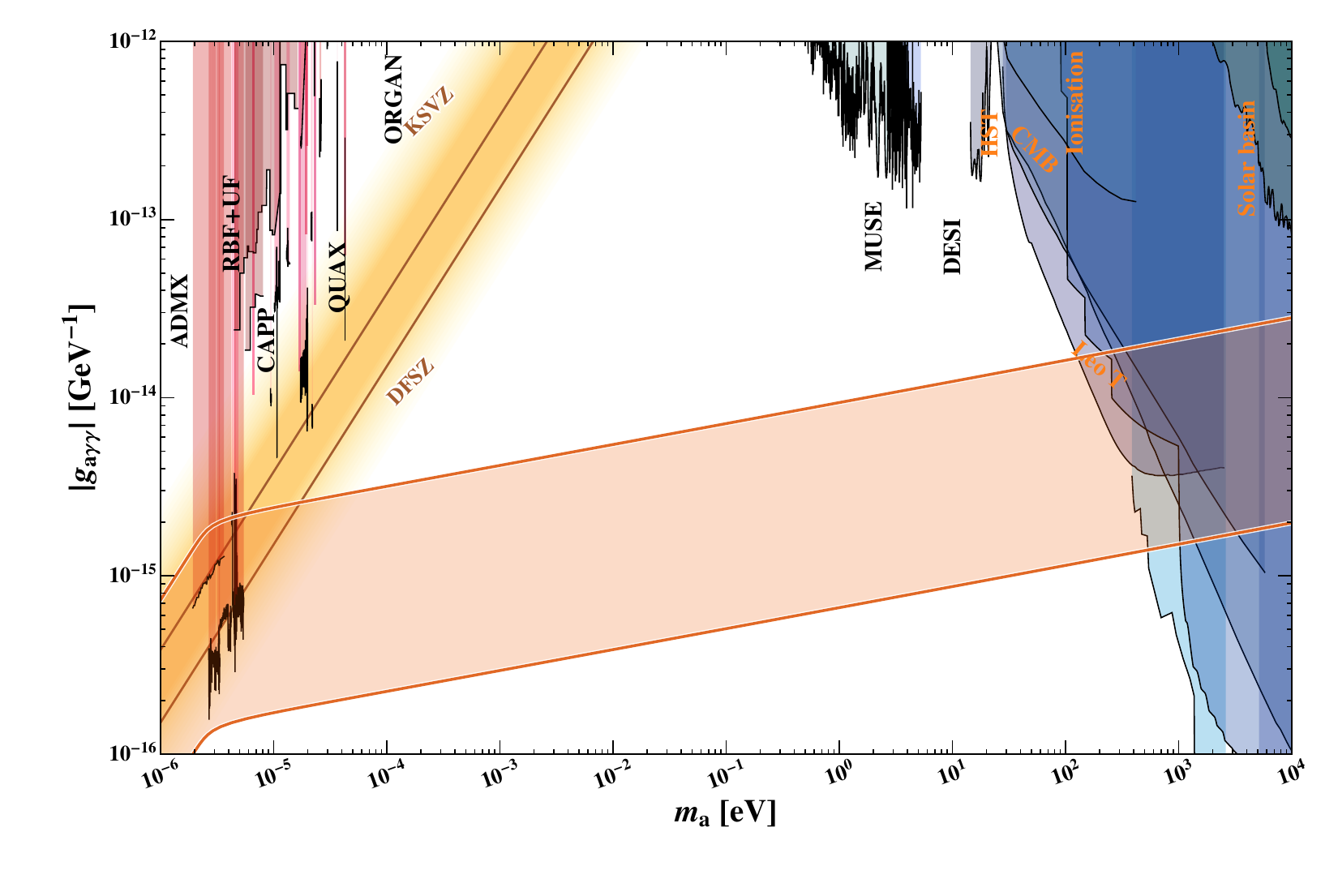}
\par\vspace{-0.6em}\small (d) $S_5$
\end{minipage}
\vspace{0.35em}

\begin{minipage}{0.49\textwidth}
\centering
\includegraphics[width=\linewidth]{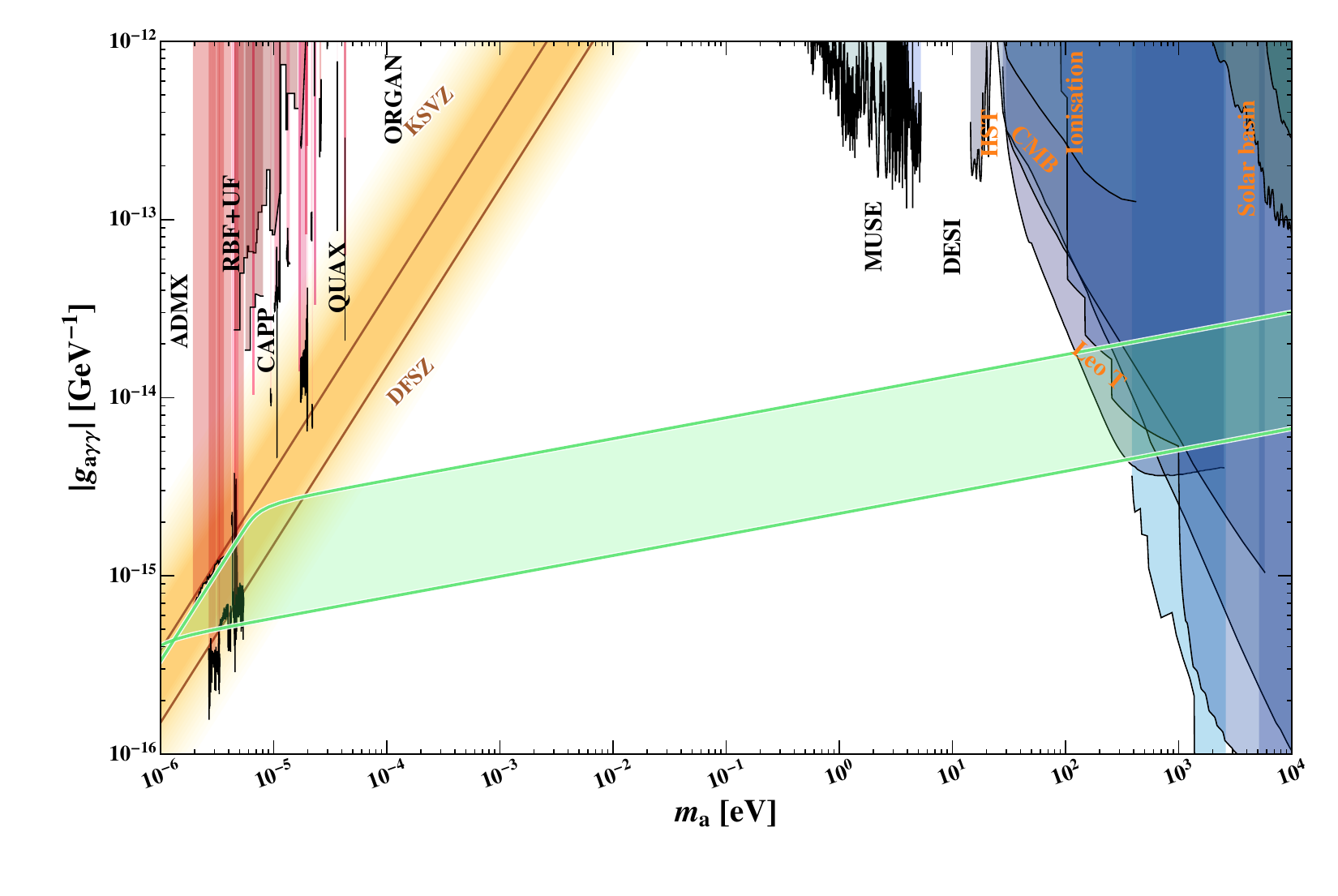}
\par\vspace{-0.6em}\small (e) $S_6$
\end{minipage}\hfill
\begin{minipage}{0.49\textwidth}
\centering
\includegraphics[width=\linewidth]{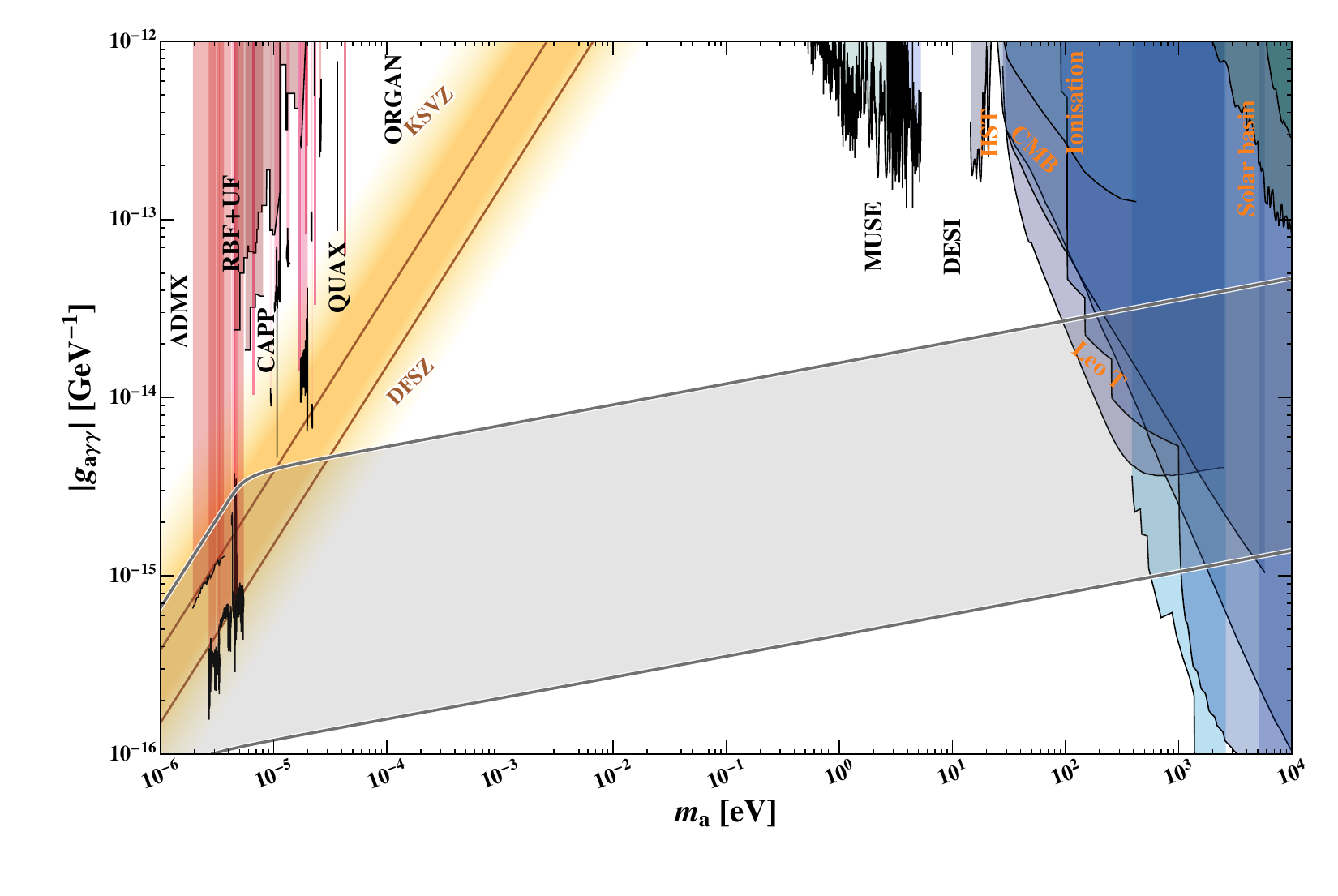}
\par\vspace{-0.6em}\small (f) $S_7$
\end{minipage}
\caption{Axion--photon predictions for the distinct-VQ sets $S_2$~(\figPatchKey{0.471,0.106,0.654}{0.884,0.803,0.924}), $S_3$~(\figPatchKey{0.952,0.736,0.088}{0.989,0.942,0.799}), $S_4$~(\figPatchKey{0.952,0.088,0.376}{0.989,0.799,0.863}), $S_5$~(\figPatchKey{0.940,0.360,0.035}{0.987,0.859,0.788}), $S_6$~(\figPatchKey{0.354,0.966,0.456}{0.858,0.993,0.880}), and $S_7$~(\figPatchKey{0.42,0.42,0.42}{0.90,0.90,0.90}), for $y_Q=1$ and $5\times10^{10}\leq\Lambda/\GeV\leq5\times10^{16}$. The two boundaries of each patch are the PQ-charge assignments in Tab.~\ref{tab:distinct-maximal-uv-sets} that minimize and maximize the axion--photon coupling. The axes and exclusion regions are the same as in Fig.~\ref{fig:axion-photon-lambda-scan}.}
\label{fig:axion-photon-distinct-limits}
\end{figure*}

 In Fig.~\ref{fig:axion-photon-lambda-scan}, the plot shows the case for the one-VQ and identical-copy spectra, whereas Fig.~\ref{fig:axion-photon-distinct-limits} depicts the distinct-VQ sets in which  $5\times10^{10}\leq\Lambda/\GeV\leq5\times10^{16}$. 
At each value of $\Lambda$, we set $M_Q=y_Q\Lambda/2=\Lambda/2$, determine the axion decay constant through $f_a=v/N_{\rm DW}=\Lambda/(\sqrt{2}\,N_{\rm DW})$, and evaluate $\chi_{\rm SI}$ for the given VQ spectrum.
Accordingly, the physical axion mass $m_a$ and $g_{a\gamma\gamma}$ are obtained from Eqs.~\eqref{eq:qcd-si-axion-masses} and \eqref{eq:axion-photon-coupling}, respectively. 
The particle content and PQ assignment are kept fixed along each trajectory, and only $\Lambda$ is varied. 
Decreasing $\Lambda$ increases the aligned small-instanton contribution and moves the scenario toward larger $m_a$. 
Equivalently, at fixed physical axion mass the coupling lies below the corresponding conventional QCD-axion line by the factor $(1+\chi_{\rm SI}/\chiQCD)^{-1/2}$. 
Moving from left to right on the axis of $m_a$, in Figs.~\ref{fig:axion-photon-lambda-scan} and \ref{fig:axion-photon-distinct-limits}, the values of $\Lambda$ decrease. 
Consequently, physical axion masses increase, which are also accompanied by lowering of $f_a$ and thereby increasing the axion--photon coupling.

Figs.~\ref{fig:axion-photon-lambda-scan} and \ref{fig:axion-photon-distinct-limits} also show the limits collected by AxionLimits~\cite{AxionLimits}. 
In the plots, the laboratory and haloscope constraints include ADMX~\cite{ADMX:2025vom}, CAPP~\cite{Bae:2024kmy}, RBF+UF~\cite{Wuensch:1989sa,Hagmann:1996qd}, QUAX~\cite{QUAX:2024fut}, and ORGAN~\cite{Quiskamp:2024oet}, while the higher-mass exclusions shown include MUSE~\cite{Todarello:2023hdk}, DESI~\cite{Wang:2023imi}, HST~\cite{Carenza:2023qxh}, CMB ionization~\cite{Cadamuro:2011fd}, the solar basin~\cite{Beaufort:2023zuj}, and Leo~T~\cite{Wadekar:2021qae}. 
The shaded regions, with various colors, on the exclusion side of these boundaries are ruled out. 
The yellow region~(\figPatchKey{0.64,0.36,0.18}{0.98,0.91,0.64}) is the conventional QCD-axion band and is obtained by setting $\chi_{\rm SI}=0$ in Eq.~\eqref{eq:axion-photon-mass-with-small-instantons} and applying the KSVZ/DFSZ bounds used by AxionLimits and it spans the anomaly ratios between $E/N=5/3$ and $44/3$~\cite{DiLuzio:2016sbl,DiLuzio:2017pfr,AxionLimits}.

Fig.~\ref{fig:axion-photon-lambda-scan} contains the one-VQ models and the maximal identical-copy spectra. 
For one VQ, $E/N$ and $N_{\rm DW}$ are given in Tab.~\ref{tab:vq-copy-number-bounds}. 
Identical copies of VQs with equal PQ signs multiply $N_{\rm DW}$ by the copy number but leave $E/N$ unchanged. 
Different contours stands for different VQs and are represented by different colors. 
For instance, $R_{11}$~(\figSolidKey{0,0.28,0.67}) is a solid one-VQ trajectory, whereas $R_2^{(11)}$~(\figLongDashKey{0.05,0.23,0.52}) is a patterned dashed identical-copy trajectory and all remaining assignments for different VQs are given in the caption of the same figure.

Several contours remain inside the yellow QCD-axion band. 
For one VQ these are $R_{11}$~(\figSolidKey{0,0.28,0.67}), $R_{12}$~(\figSolidKey{0,0.52,0.38}), $R_{13}$~(\figSolidKey{0.82,0.05,0.30}), and $R_{14}$~(\figSolidKey{0.80,0.47,0.65}); while for identical copies they are $R_2^{(11)}$~(\figLongDashKey{0.05,0.23,0.52}), $R_4^{(3)}$~(\figLongDashKey{0.73,0.19,0.02}), $R_6^{(2)}$~(\figLongDashKey{0.50,0.04,0.19}), and $R_7^{(2)}$~(\figLongDashKey{0.32,0.10,0.58}). 
In these cases the loop and mass suppressions in the zero-mode closure keep $\chi_{\rm SI}/\chiQCD\ll1$ over the range of $\Lambda$ chosen, even when the instanton-size contribution is formally UV dominated. 
These are also the cases whose curves remain below the black horizontal dashed line in Fig.~\ref{fig:mass-shift-single-identical}  and their physical axion masses therefore remain close to $m_{a,{\rm QCD}}$. 
By contrast, $R_1^{(25)}$~(\figLongDashKey{0.36,0.20,0.04}), $R_9^{(5)}$~(\figLongDashKey{0.64,0,0.48}) and $R_{13}^{(2)}$~(\figLongDashKey{0.18,0.25,0.42}) lie outside the band throughout the range of $\Lambda$, whereas $R_{16}$~(\figSolidKey{0.46,0.16,0.62}), $R_{17}$~(\figSolidKey{0.90,0.28,0}), $R_3^{(8)}$~(\figLongDashKey{0,0.42,0.46}), $R_{10}^{(4)}$~(\figLongDashKey{0.10,0.40,0.03}) and $R_{11}^{(2)}$~(\figLongDashKey{0,0.38,0.66}) cross it when the UV susceptibility becomes comparable to the QCD one. 
The contrast between $R_{13}$~(\figSolidKey{0.82,0.05,0.30}) and $R_{13}^{(2)}$~(\figLongDashKey{0.18,0.25,0.42}) is especially visible, namely, the extra copy leaves $E/N$ unchanged while enhancing $\chi_{\rm SI}$ and moves the physical mass away from the QCD one.

As evident from Fig.~\ref{fig:axion-photon-lambda-scan}, the largest coupling reached by a one-VQ scenario is $|g_{a\gamma\gamma}|=4.12\times10^{-13}\,\GeV^{-1}$ at $m_a=1.61\times10^{-3}\,{\rm eV}$ for $R_{11}$~(\figSolidKey{0,0.28,0.67}). The maximum axion-photon coupling for identical VQ copy case is $9.36\times10^{-13}\,\GeV^{-1}$ at $m_a=9.66\times10^{-4}\,{\rm eV}$ for $R_7^{(2)}$~(\figLongDashKey{0.32,0.10,0.58}).

Fig.~\ref{fig:axion-photon-distinct-limits} shows the variation of the axion--photon coupling for the distinct-VQ sets $S_2$~(\figPatchKey{0.471,0.106,0.654}{0.884,0.803,0.924}), $S_3$~(\figPatchKey{0.952,0.736,0.088}{0.989,0.942,0.799}), $S_4$~(\figPatchKey{0.952,0.088,0.376}{0.989,0.799,0.863}), $S_5$~(\figPatchKey{0.940,0.360,0.035}{0.987,0.859,0.788}), $S_6$~(\figPatchKey{0.354,0.966,0.456}{0.858,0.993,0.880}) and $S_7$~(\figPatchKey{0.42,0.42,0.42}{0.90,0.90,0.90}). Each PQ-sign assignment fixes discrete values of $E/N$ and $N_{\rm DW}$ and therefore gives a definite contour. Tab.~\ref{tab:distinct-maximal-uv-sets} lists the two assignments that minimize and maximize $N_{\rm DW}|E/N-1.92|$; they form the lower and upper boundaries of each patch at fixed $\Lambda$. The colored area between them is only a visual envelope: a generic interior point is not an additional PQ assignment and need not correspond to a physical model.

The sets $S_3,\ldots,S_7$ have the same closure $(k,r)=(12,0)$ and hence the same UV behavior of instanton induced susceptibility which fixes their dependence on $M_{\rm UV}/\Lambda$. However, their positions in the axion--photon plane and 
their determinant prefactors and the values of $E/N$ and $N_{\rm DW}$ associated with the allowed PQ-charge assignments are different. 
Consequently, spectra with the same UV sensitivity can produce patches with different widths and vertical positions. 
At fixed $\Lambda$, Eq.~\eqref{eq:axion-photon-coupling} may be written as
\be
\label{eq:photon-anomaly-scaling}
|g_{a\gamma\gamma}|\propto
\frac{|E/N-1.92|}{f_a}
=\frac{\sqrt{2}\,N_{\rm DW}|E/N-1.92|}{\Lambda}.
\ee
Thus both the distance of $E/N$ from $1.92$ and the value of $N_{\rm DW}$ determine the photon coupling. 
For the distinct-VQ sets, the largest coupling outside the shaded exclusion regions is obtained for $S_2=\{R_{11},R_{13}\}$ with $(E/N,N_{\rm DW})=(34/33,22)$, giving $\left|g_{a\gamma\gamma}\right|=3.16\times10^{-14}\,\GeV^{-1}$ at $m_a=81.54\,{\rm eV}$. 
To illustrate how a PQ-sign choice fixes the width and vertical position of a distinct-VQ envelope, we consider two limiting assignments of $S_2$~(\figPatchKey{0.471,0.106,0.654}{0.884,0.803,0.924}), which has $(N_{\rm DW},E/N)=(22,34/33)$ and $(2,14/3)$, respectively. 
The first assignment has the smaller value of $|E/N-1.92|$ but the much larger $N_{\rm DW}$, so it gives the larger photon coupling at the same PQ breaking scale.
For the remaining distinct-VQ sets, the largest allowed couplings are $\sim$ $|g_{a\gamma\gamma}|=1.36\times10^{-14}\,\GeV^{-1}$ at $m_a=1.40\times10^{2}\,\mathrm{eV}$ for $S_3$, $1.46\times10^{-14}\,\GeV^{-1}$ at $m_a=1.33\times10^{2}\,\mathrm{eV}$ for $S_4$, $1.66\times10^{-14}\,\GeV^{-1}$ at $m_a=1.23\times10^{2}\,\mathrm{eV}$ for $S_5$, $1.78\times10^{-14}\,\GeV^{-1}$ at $m_a=1.18\times10^{2}\,\mathrm{eV}$ for $S_6$, and $2.67\times10^{-14}\,\GeV^{-1}$ at $m_a=8.99\times10^{1}\,\mathrm{eV}$ for $S_7$.

Figs.~\ref{fig:axion-photon-lambda-scan} and \ref{fig:axion-photon-distinct-limits} compare the predicted $g_{a\gamma\gamma}$--$m_a$ regions with the displayed experimental and astrophysical limits. The portions outside both the standard QCD-axion band and the exclusion regions correspond to viable aligned small-instanton predictions. Their displacement from the QCD band arises because $\chi_{\rm SI}$ increases the physical axion mass, while $g_{a\gamma\gamma}$ remains determined by $f_a$ and the electromagnetic-to-color anomaly ratio, rather than from a direct modification of the axion--photon vertex. These regions therefore provide concrete targets for axion searches beyond the conventional QCD-axion band.

\section{Summary}
\label{sec:summary}

We have evaluated small-instanton contributions to the axion potential across KSVZ spectra containing a single VQ, multiple identical copies, or sets of distinct VQ representations. The VQ content determines the number of fermion zero modes that must be saturated for nonzero instanton contribution. For a closure containing $k$ scalar--Yukawa loops and $r$ mass insertions, the condition $2|N|=2k+r$ fixes the allowed contractions, while the exponent $\Delta=b_3+r-4$ determines which endpoint of the instanton-size integral dominates. To keep the instanton estimates under perturbative control up to $M_{\rm UV}$, we also imposed gauge-perturbativity constraints on every class of VQ spectrum. The strongest UV enhancements of the small-instanton susceptibility obtained from the leading zero-mode closures are
\[
\left(\frac{M_{\rm Pl}}{\Lambda}\right)^{31/3},
\quad
\left(\frac{M_{\rm Pl}}{\Lambda}\right)^{13},
\quad
\left(\frac{M_{\rm Pl}}{\Lambda}\right)^{13},
\]
for a single VQ, identical copies and distinct VQ sets, respectively. 

We then studied the consequences of such enhancement on the axion mass. 
For $y_Q=1$ and $\Lambda=5\times10^{11}\,\GeV$, the largest one-VQ mass is $m_a=3.43\,{\rm eV}$, obtained for $R_{16}$ and $R_{17}$. 
Two identical copies of $R_{13}$ give $m_a=71.2\,{\rm MeV}$, while the largest mass among the distinct spectra is $m_a=65.3\,{\rm MeV}$, obtained for the $N_{\rm DW}=22$ assignment of $S_4$ ( cf. Figs.~\ref{fig:mass-shift-single-identical} and \ref{fig:mass-shift-distinct-multiplicity}).

The same spectra were examined in the axion-mass--photon-coupling plane. At fixed $f_a$ and fixed anomaly coefficients, the small-instanton susceptibility increases the axion mass without directly changing $g_{a\gamma\gamma}$. 
The largest one-VQ coupling, as evident from the plot is $4.12\times10^{-13}\,\GeV^{-1}$ at $m_a=1.61\times10^{-3}\,{\rm eV}$ for $R_{11}$, corresponding to $9.8\%$ of the upper QCD-band value at that mass. For identical copies, $R_7^{(2)}$ reaches $9.36\times10^{-13}\,\GeV^{-1}$ at $m_a=9.66\times10^{-4}\,{\rm eV}$, or $37.3\%$ of the upper QCD-band value. The largest coupling outside the shaded exclusions is obtained for $S_2=\{R_{11},R_{13}\}$: $|g_{a\gamma\gamma}|\simeq3.16\times10^{-14}\,\GeV^{-1}$ at $m_a=81.54\,{\rm eV}$ (cf. Figs.~\ref{fig:axion-photon-lambda-scan} and \ref{fig:axion-photon-distinct-limits}).

Several factors still remain outside the scope of the present analysis. 
Renormalizable interactions between the VQs and SM fermions could connect the heavy- and SM-fermion zero modes through additional Yukawa contractions. Higher-dimensional operators could instead saturate them through different combinations of masses, scalar expectation values and ultraviolet scales. Additional PQ- or gauge-charged scalars would provide further renormalizable contractions while also modifying the gauge beta functions, instanton determinants and threshold matching. Non-degenerate VQ thresholds, the coupled running of the VQ Yukawa and scalar couplings, and relative phases between the QCD and ultraviolet potentials require a model-dependent treatment within a specified ultraviolet completion. Supersymmetric extensions of the SM and unified theories such as supersymmetric ${\rm SU}(5)$ or ${\rm SO}(10)$ provide natural frameworks in which these possibilities can be explored. Their larger scalar representations, fermionic superpartners and superpotential interactions can generate additional contractions of the instanton zero modes, reduce the number of required mass insertions and predict distinct axion--mass and axion--photon coupling.

Our results show that small-instanton effects can become significant and can enhance the axion-mass and axion-photon coupling in the KSVZ axion models. 
They further show that the resulting axion mass is controlled not by the VQ representations alone, but also  by the complete UV matter content and by the interactions responsible for saturating the instanton zero modes.

\section*{Acknowledgment}

 We thank Vaisakh Plakkot for drawing our attention to the extended catalogue of KSVZ models.
This work is partially supported by the National Natural Science Foundation of China (under Grant No. 12275140) and Nankai University.

\appendix\renewcommand{\theequation}{\Alph{section}\arabic{equation}}


\section{Group theoretical factors}
\label{sec:appendix}

Below, we collect the various group theoretical factors and determinant normalizations used in Sec.~\ref{sec:vq-models}. 
For an ${\rm SU}(3)_c$ representation with Dynkin labels of $(p,q)$, the dimension ${\rm dim}(p,q)$ and quadratic Casimir $C_2(p,q)$ are~\cite{Slansky:1981yr}
\be
\label{eq:su3-dimension-casimir}
{\rm dim}(p,q)
=
\frac{1}{2}(p+1)(q+1)(p+q+2)\,,
\quad
C_2(p,q)
=
\frac{p^2+q^2+pq+3p+3q}{3}\,.
\ee
With the convention $T({\bf 3})=1/2$, the Dynkin index follows from ${\rm dim}(R_3)C_2(R_3)={\rm dim}({\rm adj})T(R_3)$ as
\be
\label{eq:su3-dynkin-index}
T(R_3)
=
\frac{{\rm dim}(R_3)C_2(R_3)}{{\rm dim}({\rm adj})}
=
\frac{{\rm dim}(R_3)C_2(R_3)}{8}\,.
\ee
The relevant invariants are collected together with the determinant prefactors in Tab.~\ref{tab:su3-group-invariants}. 
The Dynkin index for an ${\rm SU}(2)_W$ representation of $R_2$ is given by
\be
\label{eq:su2-dynkin-index}
T(R_2)
=
\frac{{\rm dim}(R_2)\left[\left({\rm dim}(R_2)\right)^2-1\right]}{12}\,,
\ee
so that $T({\bf 2})=1/2$ and $T({\bf 3})=2$ for the ${\rm SU}(2)_W$. 

In the ${\rm SU}(5)$ GUT-normalized hypercharge convention, the contributions to one-loop coefficient from one Dirac VQ are given by
\beqa
\label{eq:vq-one-loop-gauge-contributions}
\delta b_1
&=&
\frac{4}{5}\,{\rm dim}(R_3){\rm dim}(R_2)Y^2
\,,\quad
\delta b_2
=
\frac{4}{3}\,{\rm dim}(R_3)T(R_2)
\,,\quad
\delta b_3
=
\frac{4}{3}\,{\rm dim}(R_2)T(R_3)
=
\frac{4}{3}|N|\,.\nl
\eeqa
\begin{table*}[t!]
\centering
\normalsize
\setlength{\tabcolsep}{9pt}
\begin{tabular}{c c c c c c c c}
\hline\hline

$(R_3\,,R_2)$ & $(p,q)$ & ${\rm dim}(R_3)$ & $C_2(R_3)$ & $T(R_3)$ & $\delta b_3$ & $A(R_3)$ & $C_3(R_3\,,R_2)$ \\
\hline
$(3,1)$ & $(1,0)$ & $3$ & $4/3$ & $1/2$ & $2/3$ & $0.15$ & $1.17\times10^{-2}$ \\
$(3,2)$ & $(1,0)$ & $3$ & $4/3$ & $1/2$ & $4/3$ & $0.15$ & $1.57\times10^{-2}$ \\
$(3,3)$ & $(1,0)$ & $3$ & $4/3$ & $1/2$ & $2$ & $0.15$ & $2.10\times10^{-2}$ \\
\hline
$(6,1)$ & $(2,0)$ & $6$ & $10/3$ & $5/2$ & $10/3$ & $0.59$ & $2.84\times10^{-2}$ \\
$(6,2)$ & $(2,0)$ & $6$ & $10/3$ & $5/2$ & $20/3$ & $0.59$ & $9.22\times10^{-2}$ \\
\hline
$(8,1)$ & $(1,1)$ & $8$ & $3$ & $3$ & $4$ & $0.74$ & $3.80\times10^{-2}$ \\
$(8,2)$ & $(1,1)$ & $8$ & $3$ & $3$ & $8$ & $0.74$ & $1.65\times10^{-1}$ \\
\hline
$(10,1)$ & $(3,0)$ & $10$ & $6$ & $15/2$ & $10$ & $1.44$ & $1.57\times10^{-1}$ \\
$(15,1)$ & $(2,1)$ & $15$ & $16/3$ & $10$ & $40/3$ & $2.03$ & $5.08\times10^{-1}$ \\
\hline

$(15',1)$ & $(4,0)$ & $15$ & $28/3$ & $35/2$ & $70/3$ & $2.75$ & $2.14$ \\
$(24,1)$ & $(3,1)$ & $24$ & $25/3$ & $25$ & $100/3$ & $4.19$ & $3.83\times10^{1}$ \\
$(27,1)$ & $(2,2)$ & $27$ & $8$ & $27$ & $36$ & $4.64$ & $9.30\times10^{1}$ \\
\hline\hline
\end{tabular}
\caption{${\rm SU}(3)_c$ invariants, one-loop color contributions, and one-instanton determinant prefactors. The Dynkin labels $(p,q)$ refer to ${\rm SU}(3)_c$, and the entries are independent of $Y$.}
\label{tab:su3-group-invariants}
\end{table*}

We next provide the determinant normalization appearing in the instanton estimates for various representations of Tab.~\ref{tab:vq-copy-number-bounds}, which is given by the following expression~\cite{tHooft:1976snw}
\be
\label{eq:thooft-determinant-prefactor}
C_{N_c}
=
\frac{K_1e^{-K_2N_c}}
{(N_c-1)!(N_c-2)!}
\exp\!\left[
-\sum_t\bigl(S(t)-F(t)\bigr)\alpha(t)
\right]\,,
\ee
where, $S(t)$ and $F(t)$ count the embedded scalar and Weyl-fermion multiplicities of isospin-$t$, and $\alpha(t)$ is the corresponding finite determinant coefficient. 
We use $K_1=0.47$, $K_2=1.68$, and
\be
\label{eq:determinant-alpha-values}
\alpha(1/2)=0.146\,,
\quad
\alpha(1)=0.443\,,
\quad
\alpha(3/2)=0.853\,,
\quad
\alpha(2)=1.308\,.
\ee
 Eq.~\eqref{eq:thooft-determinant-prefactor} gives $C_3^{(0)}=1.52\times10^{-3}$ for ${\rm SU}(3)_c$ without fermions. 
 The six SM Dirac quarks then change this to the following
\be
\label{eq:sm-instanton-prefactor}
C_3^{\rm SM}
=
C_3^{(0)}\exp[12\alpha(1/2)]
=
8.76\times10^{-3}\,. 
\ee

Each color representation of the VQ is decomposed under the ${\rm SU}(2)$ subgroup containing the instanton. 
By writing $A(R_3)=\sum_t n_t\alpha(t)$, with $n_t$ denoting the multiplicity of each ${\rm SU}(2)$ representation in this decomposition, some of the ${\rm SU}(3)_c$ representations are decomposed as follows
\be
\label{eq:su2-instanton-embedding}
\begin{array}{rclcl}
3 &\rightarrow& \frac{1}{2}\oplus0\,, &\quad& A(3)=0.15\,, \\
6 &\rightarrow& 1\oplus\frac{1}{2}\oplus0\,, && A(6)=0.59\,, \\
8 &\rightarrow& 1\oplus\frac{1}{2}\oplus\frac{1}{2}\oplus0\,, && A(8)=0.74\,, \\
10 &\rightarrow& \frac{3}{2}\oplus1\oplus\frac{1}{2}\oplus0\,, && A(10)=1.44\,, \\
15 &\rightarrow& \frac{3}{2}\oplus1\oplus1\oplus\frac{1}{2}\oplus\frac{1}{2}\oplus0\,, && A(15)=2.03\,, \\
15' &\rightarrow& 2\oplus\frac{3}{2}\oplus1\oplus\frac{1}{2}\oplus0\,, && A(15')=2.75\,, \\
24 &\rightarrow& 2\oplus2\!\left(\frac{3}{2}\right)\oplus2(1)\oplus2\!\left(\frac{1}{2}\right)\oplus0\,, && A(24)=4.19\,, \\
27 &\rightarrow& 2\oplus2\!\left(\frac{3}{2}\right)\oplus3(1)\oplus2\!\left(\frac{1}{2}\right)\oplus0\,, && A(27)=4.64\,.
\end{array}
\ee
The fermion and its conjugate give equal determinant factors. 
For a vector-like pair with $R_2$ being the ${\rm SU}(2)_W$ representation, the expression of $C_3$ becomes
\be
\label{eq:vq-instanton-prefactor}
C_3(R_3\,,R_2)
=
C_3^{\rm SM}
\exp\!\left[2{\rm dim}(R_2)A(R_3)\right]\,.
\ee

\bibliography{references}

\end{document}